\documentclass{article}

\usepackage[english]{babel}
\usepackage[letterpaper,top=2cm,bottom=2cm,left=3cm,right=3cm,marginparwidth=1.75cm]{geometry}

\usepackage{amsmath}
\usepackage{amssymb}
\usepackage{mathtools}
\usepackage{graphicx}
\usepackage[colorlinks=true, allcolors=blue]{hyperref}
\usepackage{setspace} 

\usepackage{authblk}
\usepackage{comment,xcolor}
\usepackage{paralist, tabularx}

\title{Eco-evolutionary dynamics of cooperative antimicrobial resistance in a population of fluctuating volume and size}
\author{Llu\'is Hern\'andez-Navarro*, Matthew Asker, and Mauro Mobilia$\dagger$}
\affil{Department of Applied Mathematics, School of Mathematics\\ University of
Leeds, Leeds LS2 9JT, U.K.}
\affil{*L.Hernandez-Navarro@leeds.ac.uk}
\affil{$\dagger$ M.Mobilia@leeds.ac.uk}

\begin{document}
\maketitle

\begin{abstract}
Antimicrobial resistance to drugs (AMR), a global threat to human and animal health, is often regarded as resulting from cooperative behaviour. Moreover, microbes generally evolve in volatile environments that, together with demographic fluctuations (birth and death events), drastically alter population size and strain survival. Motivated by the need to better understand the evolution of AMR, we study a population of time-varying size consisting of two competing strains, one drug-resistant and one drug-sensitive, subject to demographic and environmental variability. This is modelled by a binary carrying capacity randomly switching between mild and harsh environmental conditions, and driving the fluctuating volume (total amount of nutrients and antimicrobials at fixed concentration), and thus the size of the community (number of resistant and sensitive cells). We assume that AMR is a shared public good when the concentration of resistant cells exceeds a fixed {\it concentration cooperation threshold}, above which the sensitive strain has a growth advantage, whereas resistant cells dominate below it. Using computational means, and devising an analytical treatment (built on suitable quenched and annealed averaging procedures), we fully characterise the influence of fluctuations on the eco-evolutionary dynamics of AMR, and notably obtain specific strain fixation and long-lasting coexistence probabilities as a function of the environmental variation rate and cooperation threshold. We find that microbial strains tend to coexistence, but demographic fluctuations eventually lead to the extinction of resistant or sensitive cells for small or large values of the concentration cooperation threshold, respectively. This also holds for dynamic environments, whose specific properties determine the extinction timescale.
\end{abstract}

\doublespacing

\section{Introduction}\label{Sec:Intr}
The evolution of natural populations is often strongly influenced by environmental variability (EV), characterised by endlessly changing  conditions, such as temperature, light, pH, toxins, and nutrients~\cite{Chesson81,Chesson94,vasi_1994,Chesson00,Ellner19,Murugan21}. In particular, microbial populations generally evolve in volatile environments, subject to conditions fluctuating between harsh and mild. For instance, the abundance of nutrients in a community can undergo cycles of feast and famine and the amount of toxin can suddenly and radically change~\cite{vasi_1994,Srinivasan98,proft2009,Himeoka19,Merritt18,wienand2017evolution,wienand2018eco,AHNRM23}. Exogenous changes thus cause environmental fluctuations that shape the population evolution~\cite{Kussel05,acar2008,Lambert2014,caporaso_2011,abdul2021fluctuating,rescan_2020}, and in particular the ability of species to coexist~\cite{Chesson81,Chesson94,Chesson00,Chisholm14,kalyuzhny_2015,Mitri16,meyer_2020,meyer_2021,Grilli20,West22,AHNRM23} and to act cooperatively~\cite{Brockhurst07,Brockhurst07b,Patwas09,Wahl02,sanchez2013feedback,wienand2018eco,HNARM23}.  Demographic noise (DN) stemming from  random birth and death events is another source of fluctuations. These are  negligible in large communities and strong in small communities, where DN  can lead to the {\it fixation} of one species, when it takes over the entire population, and the extinction of the others~\cite{Ewens,Kimura,Blythe07,Plotkin23}. The  dynamics of the make-up and size of a population are often coupled~\cite{Roughgarden79}, resulting  in its eco-evolutionary dynamics~\cite{Pelletier09,Harrington14,wienand2018eco,taitelbaum2020population,taitelbaum2023evolutionary}. The joint effect of EV and DN is crucial in microbial communities where it can lead to population bottlenecks, resulting in colonies prone to fluctuations~\cite{Wahl02,Patwas09,Brockhurst07,Brockhurst07b}, and hence can greatly influence the eco-evolutionary dynamics of antimicrobial resistance (AMR)~\cite{Coates18,marrec2020resist,AHNRM23}. It is also worth mentioning the considerable recent efforts dedicated to researching the mechanisms underpinning the coexistence of competing species under various scenarios, see e.g. \cite{Leibold19,Pinsky19,Shibasaki2021,Gore22}. Moreover, the influence of different kinds of variability (e.g. quenched disorder, heterogeneous rates) on species coexistence has notably been investigated in ecosystems~exhibiting cyclic dominance, see e.g. \cite{Szolnoki14,Dobra18} and references therein.

AMR, whose rise is a major societal threat~\cite{oneill2016tackling,Lancet2022}, can often be interpreted as resulting from a cooperative behaviour. This occurs when antimicrobial drugs are inactivated by an enzyme produced, at a metabolic cost, by resistant cells~\cite{davies1994inactivation,Wright05,Yurtsev13,vega2014collective,Bottery2021,AHNRM23}. When the concentration of resistant microbes exceeds a fixed ``concentration cooperation threshold'' there are enough drug-inactivating enzymes for the protection against the toxin to be shared with sensitive cells at no metabolic cost. However, when the concentration of resistant cells is below the cooperation threshold, only resistant microbes benefit from the protection of the enzymes. In this scenario, AMR is mediated by drug-inactivating enzymes that act as a public good only above the cooperation threshold. This yields the spread of the resistant strain below the threshold, where the drug limits the growth of (or kills) sensitive microbes, while these thrive above the cooperative threshold where the drug-inactivating enzyme concentration is high~\cite{davies1994inactivation,Wright05,Yurtsev13,AHNRM23}. In a large population and static environment, sensitive and resistant strains thus coexist~\cite{Yurtsev13,vega2014collective, meredith2015collective, bottery2016selective}. However, this picture can be greatly altered by fluctuations arising in finite population subject to EV~\cite{Wahl02,Brockhurst07,Brockhurst07b,Patwas09,shade2012,stegen2012,Coates18,AHNRM23}.
 
Motivated by the need to better understand the evolution of AMR, we study the influence of EV and DN on the eco-evolutionary dynamics of a population of time-varying size consisting of two competing species, an antimicrobial-resistant strain and another sensitive to antimicrobial drugs. Here, EV is modelled by a binary carrying capacity that randomly switches between states corresponding to mild and harsh conditions (high and low values, respectively), with the antimicrobial drug's concentration kept constant. This choice of EV, that allows us to model population bottlenecks, is responsible for the fluctuating size of the community. This encompasses the time-variation of the number of cells of each strain, and the amount of nutrients and toxins in  the community at fixed concentration (varying volume). Hence, this EV mimics naturalistic settings subject to sudden floods/drainage events that greatly alter their volume (such as in rivers, sewerage, and natural ponds) in the context of antimicrobial-polluted environments~\cite{gothwal2015antibiotic}, as well as in microdroplet chemostat lab setups of time-varying volume~\cite{jakiela2013bacterial,totlani2020scalable}. Complementarily, in references~\cite{HNARM23,AHNRM23} EV was modelled by a binary carrying capacity switching  at constant volume, i.e. a form of EV with time-varying concentration of nutrients and toxins. In what follows, we discuss the microbial behaviour under volume-fluctuating environments and, by combining analytical and computational means, we determine the environmental conditions for the long-lived coexistence of the species and the fixation properties of each strain. We determine the fixation-coexistence diagrams of the model, and find the nontrivial environmental conditions separating the phases of dominance/fixation and long-lived coexistence of the species. We rationalise our findings by devising an analytical approach built on suitable quenched and annealed averaging procedures, that allow us to capture the fixation-coexistence diagrams, and to obtain the conditional probability that resistant cells prevail as a function of the environmental variation rate and cooperation threshold. 

The organisation of the paper is as follows: in the next section, we introduce the model and our general methods. Section~\ref{Sec:StaticEnv} is dedicated to the discussion of the model properties in a static environment, first in an infinitely large population (mean-field analysis) and then in a finite community. In section~\ref{Sec:DynEnv}, we present the results of the eco-evolutionary dynamics in a fluctuating environment, by focusing on the fixation-coexistence diagrams and their analysis in sections~\ref{Sec:DynEnv:Theory} and \ref{Sec:DynEnv:FixAndCoexDiagram}. Our conclusions are presented in section~\ref{Sec:Conclusion}, and some additional technical details can be found in the appendices.

\section{Model \& methods}\label{Sec:MethMod}
In this section, we introduce the idealised microbial community model whose eco-evolutionary dynamics we study, and describe the modelling methods used in our analysis.

\subsection{Model}\label{Sec:MethMod:Model}
We consider a well-mixed population of time-varying size $N(t)=N_R(t) + N_S(t)$ consisting of $N_R$ resistant ($R$) microbes and $N_S$ sensitive ($S$) microbes, which compete for the same resources in the presence of a constant influx of antimicrobial drug. The strain $R$ experiences a fixed metabolic cost (reduced reproduction rate) to constantly produce resistance enzymes\footnote{As opposed to the case where \(R\) can regulate the production of resistance enzyme by quorum sensing~\cite{pai2012optimality,zhao2019behavioral}. The influence of this mechanism on the evolution of AMR is not addressed in this work.} that break down the drug in their immediate vicinity, protecting them and also reducing the overall drug concentration in the community. Microbes of strain $S$ pay no metabolic cost associated to drug-inactivating enzymes (of which they are not producers), but their growth is hampered by the presence of the toxin. When the fraction of $R$ in the population is sufficiently high, they produce enough resistance enzymes to bring the overall drug concentration below its minimum inhibitory concentration (MIC)~\cite{davies1994inactivation,brook2004beta,Yurtsev13,bottery2016selective}. When this happens, the drug is ineffective everywhere and $S$ benefits from the enzymes protection despite not having to pay any cost for their production. The production of the resistance enzymes by $R$ making up a sufficiently high fraction of the community can hence be regarded as cooperative behaviour, and their protection against antimicrobials as a form of \emph{public good}~\cite{Yurtsev13,vega2014collective,Bottery2021}, see figure~\ref{Fig:Sketch}(a).

For concreteness, we assume a biostatic antimicrobial that only affects the growth rate of  $S$.\footnote{The case where the drug increases the death rate of the $S$ microbes corresponds to a biocidal antimicrobial, and is not directly considered here. This is a stringent limitation, since a given drug may act as a biocidal or biostatic antimicrobial depending on its concentration~\cite{sanmillan2017fitness}.} The metabolic cost for $R$ to produce the enzymes is denoted by $s$ and the growth hindrance experienced by $S$ microbes in the presence of the drug (above the MIC) is denoted by $a$. In this setting, the fraction of $R$ in the population, given by $x\equiv N_R / N$, determines whether the enzymes offer a public good protection to $S$. If $x_\text{th}$ is the fraction of $R$ necessary for the enzymes to share their protection across the community, $S$ is protected when $x\geq x_{\text{th}}$, whereas it is affected by the drug when $x<x_\text{th}$. Accordingly, the growth fitnesses $f_{R/S}$ of the $R$ and $S$ microbes are
\begin{equation}
    f_R=1-s\hspace{0.5cm}\text{and}\hspace{0.5cm}f_S=1-a\theta[x_\text{th}-x],
\end{equation}
where \(\theta[z]\) is the Heaviside step function, defined as \(\theta[z]=1\) if (\(z>0\)) and  \(\theta[z]=0\) otherwise, \(s\) is the resistance metabolic cost, and \(a\) is the drug-driven growth hindrance~\cite{andersson2007biological,hughes2012selection,sanmillan2017fitness,melnyk2015fitness,van2011novo}. The average population fitness is \(\bar{f}=f_RN_R/N+f_SN_S/N\). It is worth noting  that the concentration threshold  $x_\text{th}$ is constant in this setting, while it varied in~\cite{HNARM23} where the population was assumed to vary at constant volume.

The population thus evolves according to the multivariate birth-death process~\cite{Gardiner,VanKampen,Ewens} defined by
\begin{align}
\label{reactions}
N_{R/S}&\xrightarrow[]{T^+_{R/S}}N_{R/S}+1, \nonumber\\N_{R/S}&\xrightarrow[]{T^-_{R/S}}N_{R/S}-1,\end{align} 
with transition rates~\cite{wienand2017evolution,wienand2018eco,taitelbaum2020population,Shibasaki2021}
\begin{align}
T^+_{R}&=\frac{f_{R}}{\bar{f}}N_{R}=\frac{(1-s)x}{1-a\theta\left[x_{\text{th}}-x\right]+(a\theta\left[x_{\text{th}}-x\right]-s)x}N, \qquad  
T^-_{R}=\frac{N^2}{K}x \qquad \text{and}\qquad \nonumber\\
T^+_{S}&=\frac{f_{S}}{\bar{f}}N_{S}=\frac{(1-a\theta\left[x_{\text{th}}-x\right])(1-x)}{1-a\theta\left[x_{\text{th}}-x\right]+(a\theta\left[x_{\text{th}}-x\right]-s)x}N, \qquad
T^-_{S}=\frac{N^2}{K}(1-x),
\label{eq:transrates}
\end{align}
where the growth is limited by the logistic death rate $N/K$, and where $K$ is the carrying capacity, that is here assumed to be a time-fluctuating quantity, see below. This choice ensures that \(N\) obeys the standard logistic dynamics in the mean field limit, see equation \eqref{det_eq_N}. In the expressions of $T^+_{R/S}$, without loss of generality and for mathematical convenience, we have normalised $f_{R/S}$ by the average fitness \(\bar{f}\)~\cite{HNARM23}. The growth rate of each strain is thus given by its fitness relative to the average population's fitness, which is a common assumption for many biological and evolutionary processes~\cite{Chesson81,Ewens,traulsen2009stochastic}. This allows us to establish a neat relationship with  the classical Moran process, the reference birth-death-like process used to model the evolution of idealised populations of constant total size~\cite{Moran, Ewens, antal2006fixation, Blythe07, Cremer11}.

\subsection{Environmental fluctuations \& master equation}\label{Sec:MethMod:Env}
\begin{figure}
\centering
\includegraphics[width=1\textwidth]{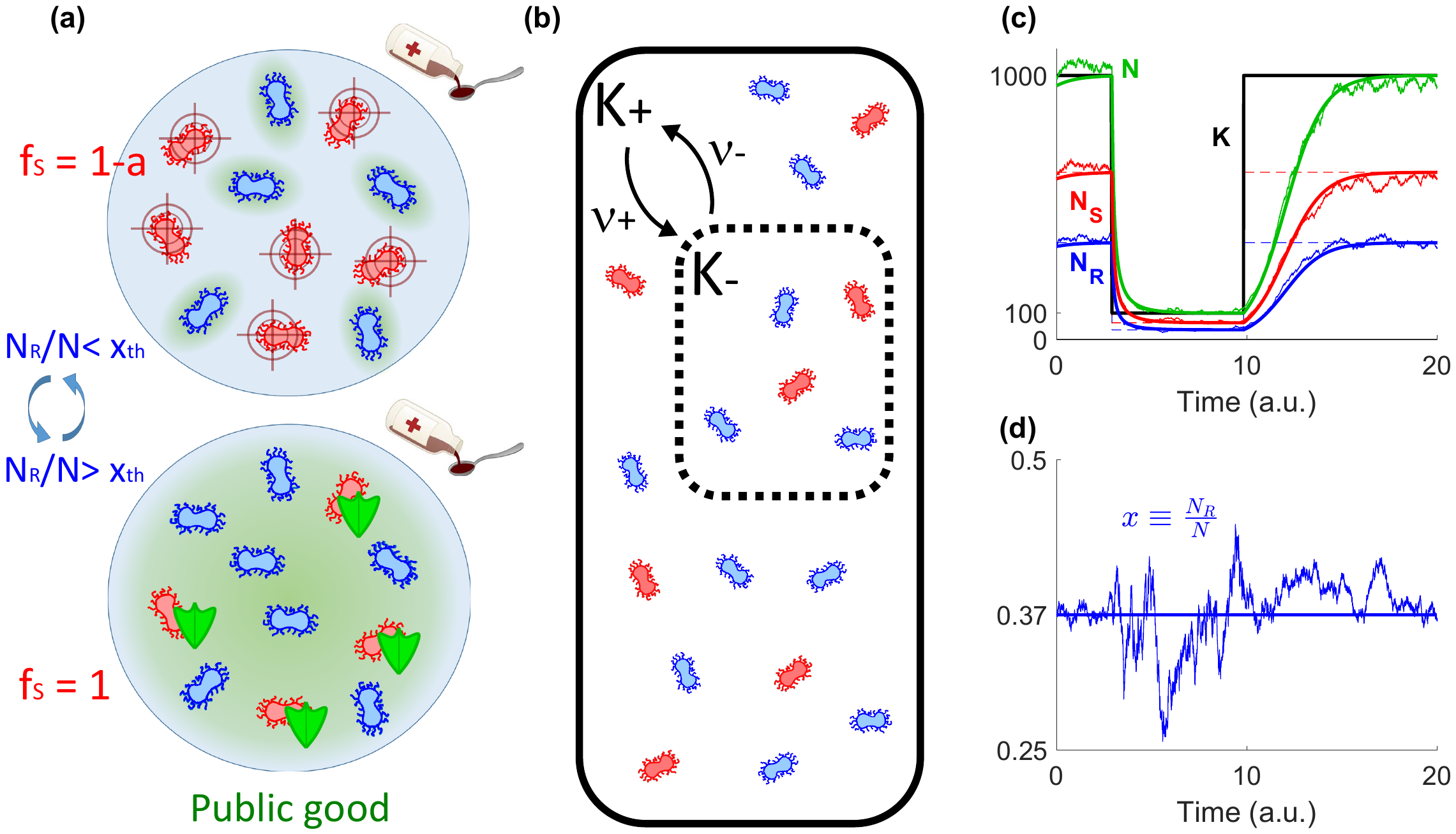}
\caption{\label{Fig:Sketch}
\textbf{Model.} {\bf (a)} Top: when the fraction of \(R\) (blue microbes) is below the concentration cooperation threshold \(x_{\text{th}}\), antimicrobial drug hinders the growth rate of \(S\) (red microbes) and \(R\) has a growth advantage. Bottom: in a cooperative scenario arising above the concentration cooperation threshold, resistance becomes shared (green shields) as the fraction of \(R\) exceeds \(x_{\text{th}}\) and these generate enough {\it resistance enzymes} (public good, green shade) to break down the drug and set its concentration below the MIC for the whole community. {\bf (b)} Ecological dynamics as random extensive changes. The environment has a constant concentration of nutrients and antimicrobial drug, but switches from high to low volume (i.e. from high to low carrying capacity) and vice-versa at rates \(\nu_{+}\) and \(\nu_{-}\), respectively. The number of microbes of each species, $N_R$ and $N_S$, evolves in this changing environment according to equation~\eqref{eq:ME}. {\bf (c)} Temporal eco-evolutionary dynamics of the microbial community for parameters \(x_{\text{th}}=0.37\), \(s=0.1\), \(a=0.25\), \(K_{-}=100\), \(K_{+}=1000\), \(\nu=0.1\), \(\delta=0.5\), and for initial conditions \(K(t=0)=K_{+}\), \(N_R(t=0)=x_{\text{th}}K_{+}\), and \(N_S(t=0)=\left(1-x_{\text{th}}\right)K_{+}\); thick black line shows the sample path of the time-switching carrying capacity \(K(t)\), thick solid coloured lines correspond to a realisation of the piecewise deterministic Markov process that ignores demographic fluctuations  and is defined by equations~\eqref{det_eq_x} and~\eqref{PDMP_N} (see section~\ref{Sec:DynEnv:FixCoexQSCD}),  for the total population (\(N\), green), number of \(R\) (\(N_R=xN\), blue), and number of \(S\) (\(N_S=(1-x)N\), red); noisy lines are the corresponding stochastic realisation of the full model under the joint effect of demographic and environmental fluctuations; dashed lines show the piecewise \hbox{(meta-)stable} equilibrium \(N_R=x_{\text{th}}K(t)\) (blue) and \(N_R=(1-x_{\text{th}})K(t)\) (red). {\bf (d)} \(R\) fraction \(x=N_R/N\) for the same sample path of varying environment as in (c); line styles as in  panel (c). See text for more details.}
\end{figure}

Here, we model EV by allowing the carrying capacity $K$ to randomly switch between two values, where this represents changes in the volume (and therefore population size) of the system. This allows us to capture sudden extreme environmental changes, as often used in laboratory experiments \cite{Wahl02,Patwas09,Brockhurst07,Brockhurst07b,Lambert2014,wienand2017evolution,HNARM23,acar2008,Lambert2014,abdul2021fluctuating,nguyen2021}, with a simple mathematical setup. Concretely, EV is driven by a dichotomous Markov noise (DMN) \cite{Bena2006,HL06,Ridolfi11}, with the transition
\begin{equation}
 \xi \longrightarrow -\xi \label{switch}
\end{equation}
occurring at rate $(1-\delta \xi)\nu$, where $\nu$ is the (average) switching  rate and $\delta$ denotes the asymmetry in the DMN switching ($-1<\delta<1$, see below), and we refer to the value of $\xi$ as the environmental state ($\xi=1$ for the mild state, and  $\xi=-1$ for the harsh). Here, the DMN is always at stationarity, implying that we have $\xi=\pm 1$ with probability $(1\pm \delta)/2$. The stationary DMN ensemble average is thus $\langle \xi(t)\rangle=\delta$ and its auto-covariance is $\langle \xi(t)\xi(t')\rangle - \langle\xi(t)\rangle\langle\xi(t')\rangle =(1-\delta^2)e^{-2\nu|t-t'|}$, which, up to a constant, coincides with the DMN auto-correlation. We note that $\nu$ is the average switching rate, and $1/(2\nu)$ is the DMN correlation time. Since we introduce the fluctuations in the environmental volume by changes in the carrying capacity, we model the binary switching carrying capacity by~\cite{wienand2017evolution, wienand2018eco, west2020, taitelbaum2020population,HNARM23,AHNRM23}
\begin{equation}
K(t)=\frac{1}{2}\left[K_+ +K_- +\xi(t)(K_+ - K_-)\right]
  \label{K(t)}.
\end{equation}
Accordingly, $K(t)$ switches between a state of high volume (\(\xi=1\)), where resources are abundant ($K_+$), to another state of low volume (\(\xi=-1\)), where they are scarce ($K_- < K_+$), with rates $\nu_+\equiv\nu(1-\delta)$ and $\nu_-\equiv\nu(1+\delta)$ according to \[K_-\xrightleftharpoons[\nu_+]{\nu_-}K_+.\]  This reflects the fact that variations of $K(t)$ are accompanied by those of the volume (amount of nutrients and toxins) at fixed concentration, see figure~\ref{Fig:Sketch}(b).

Then, EV can be characterised by the mean switching rate \(\nu\equiv(\nu_-+\nu_+)/2\) and the environmental switching bias \(\delta\equiv(\nu_--\nu_+)/(\nu_-+\nu_+)\), where $\delta>0$ corresponds to $\xi=1$ being more likely than  $\xi=-1$ (and hence more time spent, on average, in the environmental state with a high carrying capacity $K(t)=K_+$), with $\delta=0$ for symmetric DMN switching. The time-fluctuating carrying capacity \eqref{K(t)} modelling {\it environmental fluctuations} captures the time-variation of the population volume and  size, and is coupled with the birth-and-death process defined by equations~\eqref{reactions} and~\eqref{eq:transrates}.

Therefore, the master equation (ME) for the probability $P(N_R,N_S,\xi,t)$ that at time $t$ the population consists of $N_R$ and $N_S$ microbes and is in the environmental state $\xi$  is~\cite{Gardiner}:
\begin{align}
\label{eq:ME}
\hspace{-5mm}
\frac{\partial P(N_R,N_S,\xi,t)}{\partial t} &=  \left( \mathbb{E}_R^--1\right)\left[T^+_R P(N_R,N_S,\xi,t)\right]+\left( \mathbb{E}_S^--1\right)\left[T^+_S P(N_R,N_S,\xi,t)\right] \nonumber \\
&+
\left( \mathbb{E}_R^+-1\right)\left[T^-_R P(N_R,N_S,\xi,t)\right]  +\left( \mathbb{E}_S^+-1\right)\left[T^-_S P(N_R,N_S,\xi,t)\right] \\
&+ \nu_{-\xi} P(N_R,N_S,-\xi,t)-\nu_\xi P(N_R,N_S,\xi,t), \nonumber
\end{align}
where \(\mathbb{E}^{\pm}_{R/S}\) are shift operators such that \(\mathbb{E}^{\pm}_{R/S} f(N_{R/S},N_{S/R},t)=f(N_{R/S}\pm 1,N_{S/R},t)\), and the probabilities are set to \(P(N_R,N_S,\xi,t)=0\) whenever \(N_R<0\) or \(N_S<0\). The last line on the right-hand-side of equation~\eqref{eq:ME} stems from random environmental switching and is thus responsible for the coupling of EV and DN.

Since $T^{\pm}_{R/S}=0$ when $N_{R/S}=0$, there is extinction of $R$ and fixation of $S$ (\(N_S=N\)), or fixation of $R$ (\(N_R=N\)) and extinction of $S$. Here, the fixation of one strain is therefore accompanied by the extinction of the other, and when this occurs the population composition no longer varies (even though its size continues to fluctuate)\footnote{The final state of this is the absorbing state $N_R=N_S=0$, which corresponds to the eventual collapse and  extinction of the population. This occurs after a time growing exponentially with the size of the system~\cite{Spalding17, wienand2017evolution, wienand2018eco, taitelbaum2020population}. Here, where $K(t)\gg 1$, this phenomenon is practically unobservable and will therefore not be considered.}. The multivariate ME \eqref{eq:ME} can be simulated exactly using standard stochastic methods~\cite{Gillespie76,Gibson00,Anderson07,HNARM23}, and encodes the eco-evolutionary dynamics of the model. In all our simulations, $\xi$ is at stationarity, and we thus  initially have $\xi(0)=\pm1$ and hence $K(0)=K_{\pm}$ with probability $(1\pm \delta)/2$, see appendix section~\ref{Supp:Sec:sim}. For convenience and without loss of generality, the initial population size is chosen to coincide with the value of the carrying capacity at $t=0$, i.e.  $N(0)=K(0)$. We have checked that this choice has no influence on the quasi-stationary distribution of $N$, and therefore on the discussion that follows.

\section{Static environment}\label{Sec:StaticEnv}
To gain an insight into the model's eco-evolutionary dynamics, it is instructive to analyse its properties in a static environment. In this section, we thus consider that there is no EV, with the carrying capacity kept constant, $K(t)=K_0$. Below we show that the proportion of resistant microbes always tends to the fixed cooperation threshold \(x_{\text{th}}\) in static environments. However, when the microbial population is scarce, fluctuations due to birth-death events are significant, and demographic noise can thus be responsible for the extinction or fixation of the resistant strain.

\subsection{Mean-field: The absence of fluctuations promotes coexistence}
\label{Sec:StaticEnv:MeanField}

It is useful to start with the case of a very large population and carrying capacity \(K(t)=K_0\gg 1\). In this mean-field setting, we ignore demographic and environmental fluctuations, and the population's dynamics is aptly described by the mean-field (deterministic) differential equations~\cite{wienand2017evolution, wienand2018eco} 
\begin{equation}
    \dot{N}=\sum_{\alpha=R,S}\left(T^+_{\alpha}-T^-_{\alpha}\right)=N\left(1-\frac{N}{K_0}\right), \label{det_eq_N}
\end{equation}
and
\begin{equation}
    \dot{x}=\frac{d}{dt}\frac{N_R}{N}=\frac{T^+_{R}-T^-_{R}}{N}-x\frac{\dot{N}}{N}=\frac{\left(a\theta\left[x_{\text{th}}-x\right]-s\right)x(1-x)}{\left(1-a\theta\left[x_{\text{th}}-x\right]\right)+\left(a\theta\left[x_{\text{th}}-x\right]-s\right)x},
    \label{det_eq_x}
\end{equation}
where the dot indicates the time derivative. It is clear from~\eqref{det_eq_N} and~\eqref{det_eq_x} that the dynamics of $N$ and $x$ are decoupled at the mean-field level. This stems from having a constant concentration cooperation threshold $x_{\text{th}}$. According to the logistic equation~\eqref{det_eq_N}, the population size reaches its equilibrium, \(N=K_0\), on a time scale \(t\sim \mathcal{O}(1)\), while equation~\eqref{det_eq_x} implies that the population composition has a stable equilibrium \(x=x_{\text{th}}\) that is reached on a time scale  \(t\sim \mathcal{O}(1/s)\) when \(x\geq x_{\text{th}}\) and \(t \sim \mathcal{O}(1/(a-s))\) when \(x<x_{\text{th}}\). Moreover, when \(s<a\ll1\), the metabolic cost $s$ and growth hindrance $a$ are small and there is a timescale separation: \(N\) relaxes to $K_0$ much faster than \(x\) reaches  $x_{\text{th}}$. We note that the coexistence equilibrium in terms of \(R\) and \(S\) is $N_{R}=x_{\text{th}}K_0$ and $N_{S}=(1-x_{\text{th}})K_0$. This stable coexistence arises from sensitive microbes out-competing resistant ones in the presence of antimicrobial drug, but only if the resistance is shared, which requires a minimum fraction of resistant cells in the population.

\subsection{The fate of a large population in a static environment is determined by the cooperation threshold}\label{Sec:StaticEnv:FixAndCoex}
We now consider that the population size is large but finite, and the environment is static, with a constant carrying capacity $K(t)=K_0$. In this setting, the long-time dynamics is characterised by the eventual fixation of one of the strains due to birth-death demographic fluctuations. Below, we show that for large concentration cooperation thresholds (\(0\ll x_{\text{th}}<1\)) the resistant strain fixates the population quickly, whereas sensitive cells rapidly take over under smaller thresholds (\(0<x_{\text{th}}\ll1\)). Moreover, we find the cooperation threshold \(x_{\text{th}}^{*}\) at which both strains have the same fixation probability, and obtain a simple expression for the coexistence duration that increases exponentially with the total population size.

As discussed in the previous section, in a large population the community composition rapidly approaches the coexistence equilibrium, which coincides with the cooperation concentration threshold, i.e., \(x\rightarrow x_{\text{th}}\), see equation~\eqref{det_eq_x}, and the total population reaches its steady state \(N\to K_0\) before any fixation/extinction events likely occur, see equation~\eqref{det_eq_N}. We can therefore analyse the fate of a large finite population in a static environment by assuming a constant total population size coinciding with the carrying capacity, \(N=K_0\), and that any fixation/extinction events occur from the population coexistence equilibrium $x_{\text{th}}$.

When the fixed total population $N\approx K_0$ is neither too small nor too large (\(1\ll K_{0}\sim K_{-}\ll K_{+}\) in our examples), demographic fluctuations are significant (of order $\sqrt{K_0}$) and can thus prevent long-lived coexistence by leading to the fast fixation of one strain. Here, we are interested in characterising these scenarios of dominance by one strain or coexistence. For this, the evolutionary dynamics is modelled in terms of the analytically tractable Moran process~\cite{Moran,Ewens,Blythe07,Cremer11,wienand2017evolution,wienand2018eco}. In the Moran stochastic dynamics, each birth/death of one $R$ is balanced by the simultaneous death/birth of one $S$. This ensures that the overall population size in the Moran model remains constant at each update of its composition, as $N=K_0$. The simultaneous Moran birth/death events thus occur according to the reactions \[N_R+N_S\xrightarrow[]{\widetilde{T}^+_R}(N_R+1)+(N_S-1),\] and \[N_R+N_S\xrightarrow[]{\widetilde{T}^-_R}(N_R-1)+(N_S+1),\] with the effective transition rates $\widetilde{T}^\pm_R\equiv T^\pm_RT^\mp_S/N=T^\pm_RT^\mp_S/K_0$ obtained from \eqref{eq:transrates}~\cite{wienand2017evolution,wienand2018eco}:
\begin{align}
\widetilde{T}^+_{R}&=\frac{(1-s)x(1-x)}{1-a\theta\left[x_{\text{th}}-x\right]+(a\theta\left[x_{\text{th}}-x\right]-s)x}N,\text{~~and}\nonumber\\
\widetilde{T}^-_{R}&=\frac{(1-a\theta\left[x_{\text{th}}-x\right])x(1-x)}{1-a\theta\left[x_{\text{th}}-x\right]+(a\theta\left[x_{\text{th}}-x\right]-s)x}N,
\label{eq:morantransrates}
\end{align}
where $x\equiv N_R/N$ and where we used $N=K_0$ (constant population size). Therefore, the stochastic dynamics of \(x\), when the total microbial population size is kept constant (static environment), is well described by the above one-dimensional effective birth-death reactions.

{\bf Fixation probability.} We are interested in the probability that a strain takes over the entire population starting from a given initial number of microbes, a quantity referred to as {\it fixation probability}~\cite{Ewens,Kimura,antal2006fixation,Blythe07}. Here, the fixation probability of the strain $R$ when the population initially consists of \(N_R^0\) resistant individuals (and $N_S^0=N-N_R^0$ sensitive cells) is denoted by $\phi_N\left(x_{0}\right)$, where \(x_{0}\equiv N_R^0/N\) is the initial fraction of resistant microbes, and where the subscript signifies that the expression is for a population of constant size $N$. Since the fixation of $R$ coincides with the extinction of $S$, $\phi_N\left(x_{0}\right)$ also gives the extinction probability of the strain $S$.

\begin{figure}
\centering
\includegraphics[width=1\textwidth]{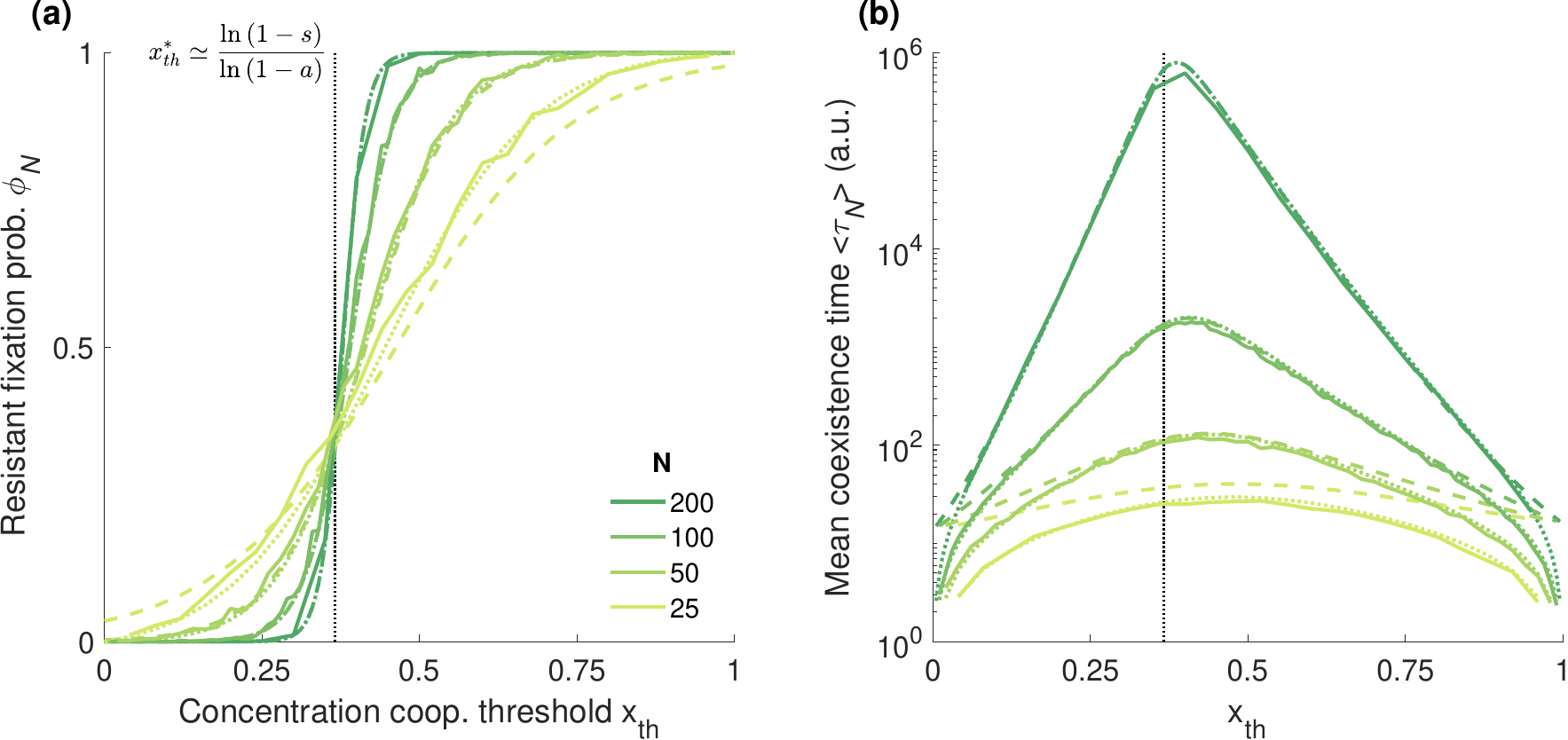}
\caption{\label{fig:FixProbAndMeanAbsTimeAndXQSD}
\textbf{\(R\) fixation probability and mean coexistence time (MCT) in static environments.} {\bf (a)} \(R\) fixation probability \(\phi_N\) in terms of the concentration cooperation threshold \(x_{\text{th}}\) for \(s=0.1\), \(a=0.25\), and for five examples of total population size, from \(N=25\) (yellow green) to 200 (dark green); the starting microbial composition is set at the coexistence equilibrium \(x_0=x_{\text{th}}\); dotted and dashed lines show the exact and approximated Moran predictions of equations~\eqref{SuppEq:FixProb} and~\eqref{eq:ApproxFixProb}, respectively, which are only distinguishable for the smallest population size; noisy solid lines are from simulation data (\(10^{3}\) realisations for each data point) and match with dotted lines. The vertical black dashed line shows  the value of \(x_{\text{th}}^{*}\) for which each strain has a probability $1/2$ to fixate when $N\gg 1$. {\bf (b)} Mean coexistence time $\langle \tau_N\rangle$ vs \(x_{\text{th}}\) in log-linear scale; the simplified formula of equation~\eqref{eq:ApproxMAT} (dashed lines), approximates well (for $\langle \tau_N\rangle > 50$) the exact MCT~\eqref{SuppEq:GenMeanAbsTime} (dotted lines) and simulation results for the MCT (solid lines); legend and line styles as in panel (a).}
\end{figure}

For the one-dimensional Moran model (e.g., two microbial strains at a fixed total population $N$), the exact general expression of the fixation probability $\phi_N\left(x_0\right)$ is well known~\cite{Gardiner, VanKampen, Ewens, antal2006fixation, traulsen2009stochastic}, see equation~\eqref{SuppEq:GenFixProb}. According to equation~\eqref{det_eq_x}, the make-up of a large population in a static environment rapidly reaches its coexistence equilibrium from any moderate initial \(R\) fraction (\(0\ll x_0\ll1\)), i.e., \(x_0\rightarrow x_{\text{th}}\), and lingers about $x_{\text{th}}$ until fixation/extinction occurs. When $x_0$ is not too close to $0$ or $1$, we can thus assume \(\phi_N(x_0)\simeq\phi_N\left(x_0=x_{\text{th}}\right)\), yielding the following approximate expression for the fixation probability of \(R\) in a population of large constant size \(N\)~\cite{HNARM23}:
\begin{equation}
    \phi_N(x_{\text{th}})\simeq\frac{1}{1+\left(\frac{1}{1-a}\right)^{-N\left(x_{\text{th}}-x_{\text{th}}^{*}\right)}}, \text{~~~with~}x_{\text{th}}^{*}\equiv \frac{\ln\left(1-s\right)}{\ln\left(1-a\right)}+\frac{1}{N}\frac{\ln\left(\frac{s(1-a)}{a-s}\right)}{\ln\left(1-a\right)};
    \label{eq:ApproxFixProb}
\end{equation}
see appendix section~\ref{Supp:Sec:ExactFixProb}, where the exact expression of $\phi_N(x_{\text{th}})$ is given by equation~\eqref{SuppEq:FixProb}. Here \(x_{\text{th}}^{*}\) is defined as the critical value of the cooperation concentration threshold for which both strains have probability \(1/2\) of fixation, i.e. $\phi_N\left(x^*_{\text{th}}\right)=1/2$. We have found that, for biologically plausible values of \(0<s<a\lesssim10^{-1}\) and \(N>25\), equation~\eqref{eq:ApproxFixProb} is a good approximation for the fixation probability and this simplified compact expression matches up well with the exact solution and simulation data; see figure~\ref{fig:FixProbAndMeanAbsTimeAndXQSD}(a). Remarkably, equation~\eqref{eq:ApproxFixProb} reveals that strain $R$ is the most likely to go extinct (and $S$ to fixate) when the cooperation threshold is below the critical value $x_{\text{th}}^*$: $\phi_N(x_{\text{th}})\ll 1$ if $x_\text{th}<x_{\text{th}}^*$. On the other hand, when the cooperation threshold exceeds $x_{\text{th}}^*$, a large concentration of $R$ is thus needed to inactivate the drug, and $R$ is likely to fixate and $S$ to go extinct, since equation~\eqref{eq:ApproxFixProb} predicts $\phi_N(x_{\text{th}})\approx 1$ if $x_{\text{th}}>x_{\text{th}}^*$. As an intriguing feature of the cooperative resistance, we notice that the more efficient in producing resistance enzymes (public good) $R$ is, the lower  $x_{\text{th}}$ is, and the more likely $R$ is to go extinct. Similarly, the less efficient in producing the public good $R$ is, the more likely it is to fixate (high $x_{\text{th}}$).

{\bf Mean coexistence time.} Another quantity of great interest is the mean time until one of the strain fixates and the other is wiped out. Since the strains here coexist until fixation/extinction of one of them, this (unconditional) mean fixation time coincides with the mean coexistence time (MCT) of the strains \(\langle \tau_N\left(x_{0}\right)\rangle\). For the Moran model in a population of fixed size $N$, the exact expression of the MCT is well-known~\cite{Gardiner, VanKampen, Ewens, antal2006fixation, traulsen2009stochastic}, see equation~\eqref{SuppEq:GenMeanAbsTime}. As discussed previously, we can assume that fixation/extinction arises from the coexistence equilibrium \(x_{0}=x_{\text{th}}\), and then obtain the simplified approximate MCT expression  
\begin{equation}
    \begin{aligned}
    \langle \tau_N(x_{\text{th}})\rangle \simeq\frac{a(1-s)}{s^2(a-s)}\frac{\phi_N(x_{\text{th}})}{x_{\text{th}}}\frac{\left(\frac{1}{1-s}\right)^{(1-x_{\text{th}})N}-1}{(1-x_{\text{th}})N},
    \end{aligned}
    \label{eq:ApproxMAT}
\end{equation}
whose derivation is given in section~\ref{Supp:SubSec:ExactMCT} of the appendix. This expression of the MCT closely reproduces the exact prediction \eqref{SuppEq:GenMeanAbsTime} and simulation data of  figure~\ref{fig:FixProbAndMeanAbsTimeAndXQSD}(b) for $\langle \tau_N\rangle >50$. We notice that $\langle \tau_N(x_{\text{th}})\rangle \ll \langle \tau_N(x_{\text{th}}^*)\rangle $ when $x_{\text{th}}$ is much below or above $x_{\text{th}}^*$. This means that, for given $s,a$ and $N$, the MCT is maximum for $x_{\text{th}}\simeq x_{\text{th}}^*$, and it decreases exponentially as $x_{\text{th}}$ deviates from \(x^*_{\text{th}}\). This can be interpreted as follows: strongly and weakly cooperative $R$ (low/high $x_{\text{th}}$, respectively) favour the rapid dominance of one strain, whereas $R$ with a cooperative threshold about $x^*_{\text{th}}$ promotes long-lived coexistence, and in this case the MCT greatly exceeds the population size~\cite{antal2006fixation,AHNRM23,AM10,AM11,cremer2009edge,he2011,reich2007}.

{\bf Coexistence probability.} In section~\ref{Sec:StaticEnv:MeanField}, we have seen that the mean-field dynamics of an infinitely large population leads to a stable coexistence of the strains. However, coexistence cannot be stable in a finite population since demographic fluctuations unavoidably cause the fixation of one stain and the extinction of the other~\cite{MA10,AM10,AHNRM23}. In fact, while we are guaranteed that only one of the strains will finally survive, its fixation can occur after a very long-lived coexistence of the strains. When this happens, the mean-field coexistence equilibrium becomes metastable. The population is thus at quasi-stationarity, and the MCT  generally scales superlinearly with the population size $N$, with a fixation/extinction time, here denoted by $\tau_N$,  that is generally exponentially distributed with a cumulative distribution  $1-{\rm exp}(-\tau_N/\langle \tau_N\rangle )$, see e.g.~\cite{MA10,AM10,AM11,assaf2017,AHNRM23}. This is in stark contrast with the MCT scaling sublinearly with $N$ when one species dominates over the other~\cite{antal2006fixation,AHNRM23,AM10,AM11,cremer2009edge,he2011,reich2007}. The scenarios of dominance and long-lived coexistence are hence  separated by a regime where the MCT scales linearly with the population size~\cite{cremer2009edge,he2011,reich2007,HNARM23,AHNRM23}. This leads us to consider that there is long-lived coexistence of $R$ and $S$ strains  whenever the coexistence time exceeds $2N$ (where the conservative factor $2$ is chosen for convenience as in~\cite{HNARM23,AHNRM23}), and otherwise one of the strains dominates and fixates in a time $\tau_N$ less than $2N$. Here, we will therefore consider that the probability to have long-lived coexistence in a population of size $N$ is $\eta_{N}={\rm Prob.}\left(\tau_N>2N\right)$, the probability that coexistence time exceeds $2N$, which is obtained from its cumulative distribution:
\begin{equation}
    \eta_{N}\equiv{\rm Prob.}\left(\tau_N>2N\right)= {\rm exp}(-2N/\langle \tau_N\rangle).
    \label{eq:StatCoexProb}
\end{equation}

Assuming that the probability of fixation of a given strain and the probability of long-lived coexistence are independent, with \eqref{eq:ApproxFixProb}, \eqref{eq:ApproxMAT} and \eqref{eq:StatCoexProb}, we thus estimate that the probabilities of fast fixation of $R$ and $S$ by time $2N$ in a population of constant size $N$ are, respectively,
\begin{equation}
\phi_N(x_{\text{th}})(1-
    \eta_{N}) \quad \text{and} \quad (1-\phi_N(x_{\text{th}}))(1-
    \eta_{N})
    \label{eq:FixbyN}.
\end{equation}
The quantities~\eqref{eq:StatCoexProb} and~\eqref{eq:FixbyN} can be used to obtain fixation-coexistence diagrams and visualise the long-time eco-evolutionary properties of the model in static environments~\cite{HNARM23,AHNRM23}. Based on these results, and assuming ${\rm min}(x_\text{th},1-x_\text{th})\lesssim 1/\sqrt{N}$ (so that demographic fluctuations about the concentration cooperation threshold are strong enough to cause fast fixation/extinction events), the long-time behaviour of a {\it finite population of size $N$ in a static environment} (with $sN\gg1$ and $(a-s)N\gg1$) can be summarised in terms of the value of $x_{\text{th}}$ relative to $x_{\text{th}}^*$. This is, if $x_{\text{th}}<x_{\text{th}}^*$, the sensitive strain $S$ dominates, while $R$ most probably fixate when $x_{\text{th}}>x_{\text{th}}^*$; whereas long-lived coexistence of the strains is expected when $x_{\text{th}}$ is about $x_{\text{th}}^*$. When ${\rm min}(x_\text{th},1-x_\text{th})\gg 1/\sqrt{N}$, demographic fluctuations are too weak to ensure fast fixation/extinction, and long-lived coexistence is always expected. The above approach and quantities \eqref{eq:StatCoexProb} and  \eqref{eq:FixbyN} are generalised to a dynamic environment in section \ref{Sec:DynEnv:Theory} to obtain the corresponding fixation-coexistence diagrams, see  equations~\eqref{eq:QFixCoexProb} and \eqref{eq:PhiA}  and figure~\ref{fig:PhaseDiagram}.

\section{Dynamic environment}\label{Sec:DynEnv}
Environments are often dynamic and endlessly change through time. Having characterised how the system evolves when the environment is static, we now look to include the biologically-motivated environmental changes driven by the binary switching of the carrying capacity~\eqref{K(t)}. These environmental switches can produce population bottlenecks, where the community size shrinks significantly and is subject to strong demographic fluctuations, hence moulding the fixation/coexistence properties of the population. In this section, we derive general analytical and numerical methods to assess the impact of environmental variability (EV) on population dynamics. In particular, we show that, in a {\it dynamic environment}, fast fixation of either strain is also possible when $x_\text{th}\approx x_\text{th}^*$, and long-lived coexistence can also occur when $x_\text{th}\approx 0$ or $1$, see section \ref{Sec:DynEnv:FixCoexQSCD}.

\subsection{Large population driven by a  switching carrying capacity: sample paths \& quasi-stationary distribution}
\label{Sec:DynEnv:Traj}
Here we show how the EV shapes the total microbial population size irrespective of composition dynamics. On the one hand we show that slow-switching environments (with small mean switching frequency \(\nu\)) yield a bimodal distribution of microbial populations, narrowly peaked about either the low (\(K_{-}\)) or high (\(K_{+}\)) carrying capacity. On the other hand, fast-switching EV (high \(\nu\)) leads to a unimodal distribution of the total population size peaked at an effective, intermediate carrying capacity \(\mathcal{K}\). The intermediate-switching regime is characterised by a mixed behaviour and a wider variability of the population distribution.

\begin{figure}
\centering
\includegraphics[width=1\textwidth]{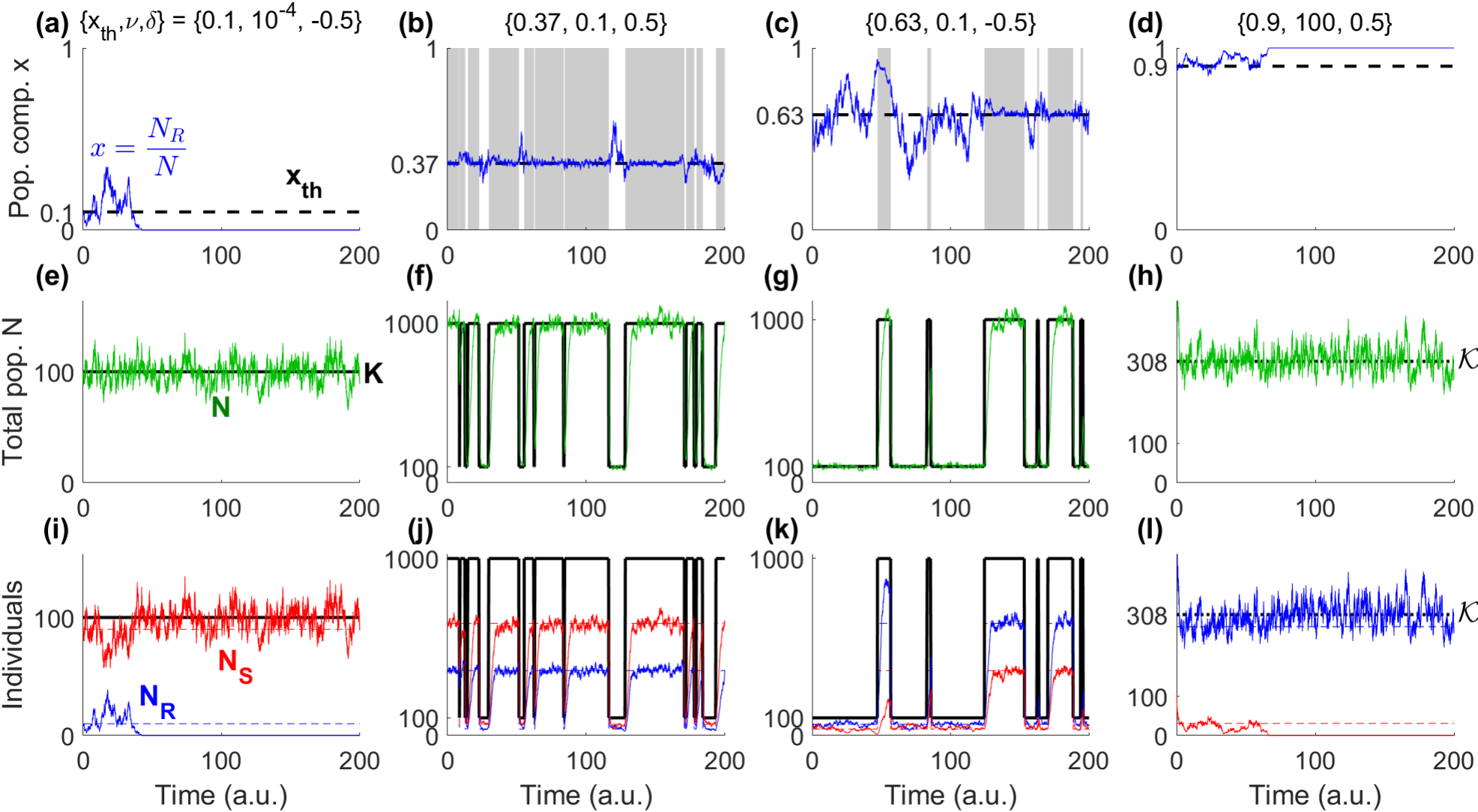}
\caption{\label{fig:Trajectories}\textbf{Eco-evolutionary dynamics sample paths.} {\bf (a)-(d)} Examples of \(R\) fraction sample paths {\it in silico} (blue lines) for four concentration cooperation thresholds \(x_{\text{th}}\) (dashed black lines), three average switching rates \(\nu\), and two environmental biases \(\delta\), for \(s=0.1\), \(a=0.25\), \(K_-=100\), and \(K_+=1000\), with initial conditions \(N_R(t=0)=x_{\text{th}}K(t=0)\) and \(N_S(t=0)=\left(1-x_{\text{th}}\right)K(t=0)\). Shaded and white areas in panels (b)-(c) encode periods of abundance (\(K(t)=K_{+}\)) and scarcity (\(K_{-}\)), respectively. Note the larger amplitude  of demographic fluctuations for the latter since  \(N\rightarrow K_{-}\ll K_+\). {\bf (e)-(h)} Same example paths as in the corresponding panels (a)-(d) for the population size \(N\) (green lines); black lines show example paths for the carrying capacity \(K\). The very high environmental switching rate \(\nu\) in panel (h) provides the effectively constant carrying capacity \(\mathcal{K}\) (black dashed line); see section~\ref{Sec:DynEnv:Traj}. {\bf (i)-(l)} Same example paths as in the previous panels for the number of \(R\) (blue lines) and \(S\) (red lines). Dashed lines show the corresponding (meta-)stable coexistence equilibrium \(N_{R}=x_{\text{th}}K\) (blue) and \(N_{S}=(1-x_{\text{th}})K\) (red); see section~\ref{Sec:StaticEnv}. The demographic noise (DN) and the low value of $x_{\text{th}}=0.1$ in panels (a),(e),(i) lead to fast extinction of \(R\), whereas the high threshold \(x_{\text{th}}=0.9\) and DN in panels (d),(h),(l) lead to an early fixation of \(R\) and extinction of \(S\).}
\end{figure}

When the population is large enough for demographic fluctuations to be negligible, and randomness only arises from the time-fluctuating environment, via the random switches of the carrying capacity~\eqref{K(t)}, the dynamics is well described by a so-called piecewise deterministic Markov process (PDMP)~\cite{PDMP}. In the PDMP, the total population dynamics between each environmental switch is deterministic and given by equation~\eqref{det_eq_N}, with $K_0$ now replaced by $K_{\pm}$ when $\xi=\pm 1$. Since equations~\eqref{det_eq_N} and~\eqref{det_eq_x} are decoupled, the PDMP does not affect the mean-field dynamics of the population composition $x$ (and vice versa), which is still given by~\eqref{det_eq_x}. Here, the PDMP for the population size $N$ (\(N\)-PDMP) is therefore defined by
\begin{equation}
    \dot{N}=N\left(1-\frac{N}{K(t)}\right)=
    \begin{cases}
    N\left(1-\frac{N}{K_-}\right), & \text{if $\xi=-1$} \\
    N\left(1-\frac{N}{K_+}\right), & \text{if $\xi=1$}
    \end{cases}
, \label{PDMP_N}
\end{equation}
where the time-varying carrying capacity \(K(t)\in \{K_-,K_+\}\) is given by equation~\eqref{K(t)}. Sample paths of this \(N\)-PDMP are shown as solid black lines in figures~\ref{Fig:Sketch}(c) and~\ref{fig:Trajectories}(e)-(h). These realisations illustrate how \(N(t)\) tracks the switching carrying capacity \(K(t)\) for low and intermediate/low switching rates $\nu$, while it fluctuates about an effective value \({\cal K}\) for very high switching rates, see  figure~\ref{fig:Trajectories}(h). The population composition \(x(t)\) always tends towards the coexistence equilibrium \(x\to x_{\text{th}}\); see figure~\ref{fig:Trajectories}(a)-(d). The dynamics of \(K(t)\) leads us to identify two limiting regimes. When environmental variability is slow, $\nu\rightarrow0$, the environmental state switches very rarely, and $N$ thus essentially coincides with $K(0)$ giving $N=K_{\pm}$ with probability $(1\pm \delta)/2$. In this slow EV regime, the population dynamics is the same as in static environments with a constant carrying capacity set by the initial condition. When the environment switches very frequently, $\nu\rightarrow\infty$, \(N\) cannot keep track of the changes of \(K(t)\), which approaches an effective value $K(t)\to {\cal K}$ obtained by self-averaging the environmental noise, yielding \(\mathcal{K}=1/\langle1/K(t)\rangle=2K_+K_-/\left[(1-\delta)K_+ + (1+\delta)K_-\right]\)~\cite{wienand2017evolution,wienand2018eco,west2020,Shibasaki2021,taitelbaum2020population,taitelbaum2023evolutionary,HNARM23,AHNRM23}. Under very high switching rate, the population size thus approaches  this effective  constant carrying capacity $N\approx {\cal K}$, see figure~\ref{fig:Trajectories}(h).

The (marginal) stationary probability density function (pdf) of the \(N\)-PDMP (ignoring demographic fluctuations) for fixed \(\{\nu,\delta\}\) is~\cite{horsthemke1984noise,HL06,Bena2006,Ridolfi11,wienand2017evolution,wienand2018eco,taitelbaum2020population,AHNRM23}
\begin{equation}
    p_{\nu,\delta}(N) = \frac{\mathcal{Z}}{N^2} \left(\frac{K_+ - N}{N}\right)^{\nu(1-\delta)-1} \left(\frac{N-K_-}{N}\right)^{\nu(1+\delta)-1},
    \label{eq:NPDMP}
\end{equation}
where \(\mathcal{Z}\) is a normalisation constant and \(N\in\left[K_-,K_+\right]\) is a continuous variable. Despite ignoring demographic fluctuations, the above equation provides a useful approximation of the actual quasi-stationary population size distribution, and captures well the underlying total population dynamics~\cite{wienand2017evolution,wienand2018eco,west2020,taitelbaum2020population,AHNRM23}. When \(\nu \ll 1\) the pdf is bimodal and  sharply peaked at \(N= K_-\) and \(N=K_+\). The probability density $p_{\nu,\delta}(N)$ begins to flatten out as \(\nu\sim1\). When the switching rate $\nu$ is increased further, the pdf becomes unimodal and sharply peaked about a value approaching $\mathcal{K}$, and eventually \(N\approx\mathcal{K}\), when \(\nu\gg1\). The actual \(N\) quasi-stationary distribution shows slightly broader peaks, due to demographic noise, but is otherwise well described by the PDMP approximation~\cite{wienand2017evolution,wienand2018eco,taitelbaum2020population,taitelbaum2023evolutionary}. In particular, the average of $N$ over the $N$-PDMP pdf given by
\begin{equation}
    \langle N\rangle_{\nu,\delta} =\int_{K_-}^{K_+} Np_{\nu,\delta}(N)~dN, 
    \label{eq:avN}
\end{equation}
is an accurate approximation of $\langle N\rangle$, the actual stationary mean population size~\cite{wienand2017evolution,wienand2018eco,taitelbaum2020population,AHNRM23}, with $\langle N\rangle\approx\langle N\rangle_{\nu,\delta}\approx [(1+\delta)K_+ +(1-\delta)K_-]/2$ when $\nu\ll 1$ and $\langle N\rangle\approx\langle N\rangle_{\nu,\delta}\approx \mathcal{K}$ when $\nu\gg 1$. As shown in references~\cite{wienand2017evolution,wienand2018eco,taitelbaum2020population,AHNRM23}, equation~\eqref{eq:avN} correctly captures that, at fixed $\delta$, the average population size $\langle N\rangle$ is a decreasing function of $\nu$.

\subsection{Insight into fixation and coexistence properties via eco-evolutionary dynamics sample paths}
\label{Sec:DynEnv:FixCoexQSCD}
It is  useful to consider the trajectories of figure~\ref{fig:Trajectories} to gain some insights into the dynamics in fluctuating environments. We first notice the strong dependence of fixation or coexistence (and which strain prevails) on $x_\text{th}$, as in static environments; see section~\ref{Sec:StaticEnv:FixAndCoex}. It is indeed clear from figure~\ref{fig:Trajectories} that, for fixation to occur in a time of order $\langle N\rangle$, $x_\text{th}$ must be sufficiently close to 0 or 1 such that demographic fluctuations are able to drive the system away from coexistence. As in section~\ref{Sec:StaticEnv:FixAndCoex}, which strain fixates is chiefly determined by whether $x_\text{th}>x_\text{th}^*$ or $x_\text{th}<x_\text{th}^*$, see figure~\ref{fig:FixProbAndMeanAbsTimeAndXQSD}(a), and fixation occurs in a finite time when ${\rm min}(x_\text{th},1-x_\text{th})\lesssim 1/\sqrt{\langle N \rangle}\approx 1/\sqrt{\langle N \rangle_{\nu,\delta}}$, see figure~\ref{fig:FixProbAndMeanAbsTimeAndXQSD}(b). Since $x_\text{th}=0.1 < x_\text{th}^*\approx 0.37$ in figure~\ref{fig:Trajectories}(a),(e),(i), it illustrates the fixation of $S$, whereas  figure~\ref{fig:Trajectories}(d),(h),(l), where $x_\text{th}=0.9>x_\text{th}^*$, exemplifies the fixation of $R$. In these two sets of panels, we also see the impact of small/large $\nu$: in figure~\ref{fig:Trajectories}(a),(e),(i) the population's initial size is $N=K_-=100$ and $\nu$ is so small that no switches occur before $S$ fixation occurs. In figure~\ref{fig:Trajectories}(d),(h),(l), $K(t)$ switches so quickly and frequently that $N\approx \mathcal{K}\approx308$, see section~\ref{Sec:DynEnv:Traj}, well before $R$ fixation occurs. In the cases where $x_\text{th}$ is close to  $x_\text{th}^*$ and far from 0 and 1, as in figure~\ref{fig:Trajectories}(b),(f),(j) and~\ref{fig:Trajectories}(c),(g),(k), we observe a long-lived coexistence of the strains.

In figure~\ref{fig:Trajectories}(b),(c), where \(x_{\text{th}}=0.37\) and \(x_{\text{th}}=0.63\) with \(\nu=10^{-1}\), we notice large demographic fluctuations when the  population experiences bottlenecks at \(N\approx K_-\) (periods in white background). We also note in figure~\ref{fig:Trajectories}(f),(g) that for an intermediate switching rate $\nu=0.1$, the population size $N$ manages to track the switches of $K$ and its distribution is expected to be well approximated by the pdf \eqref{eq:NPDMP}.

These features are essentially in line with the discussion of section~\ref{Sec:StaticEnv:FixAndCoex}, and hence similar to the behaviour found in static environments. In the next section, we will see that, in a {\it dynamic environment}, fixation of either strain is also possible when $x_\text{th}\approx x_\text{th}^*$ and that there can be long-lived coexistence also when $x_\text{th}\approx 0,1$. These are distinctive effects of EV that are analysed in detail in what follows, see figure~\ref{fig:PhaseDiagram}.

\subsection{Theory for the fixation-coexistence diagrams
in fluctuating environments}
\label{Sec:DynEnv:Theory}
In this section we develop the analytical methods for fluctuating environments allowing us to predict when, in our simulations, one of the strains will fixate or go extinct, and when both strains will coexist for long-periods. The theory that we have devised and discuss here is used to analytically reproduce, see figure~\ref{fig:PhaseDiagram}(e)-(h), the fixation-coexistence diagrams of figure~\ref{fig:PhaseDiagram}(a)-(d) obtained from extensive computer simulations. Here, we show that microbial populations in time-varying environments can exhibit two qualitatively distinct behavioural regimes, depending on whether the environment has switched at least once. If this is the case, microbes experience a mixture of the two possible environments, and dynamics is effectively captured by averaging fixation/extinction probability fluxes over the total population distribution. When switches are unlikely to occur, the  environment is essentially static and the population can be assumed to coincide with the initial value of the carrying capacity.

We first define {\it long-lived coexistence} in fluctuating environments by direct extension of what has been discussed in static environments; see section~\ref{Sec:StaticEnv:FixAndCoex}. For given EV defined by the environmental mean switching rate and bias parameters \(\{\nu,\delta\}\), there is long-lived coexistence of the strain $R$ and $S$ if the  fixation time, here denoted by $\tau$, exceeds $2\langle N\rangle$~\cite{HNARM23,AHNRM23}. Note that the mean fixation/extinction time, denoted by $\langle \tau\rangle$, coincides with the {\it mean coexistence time}, as in the environment-static case of section~\ref{Sec:StaticEnv:FixAndCoex}. We will thus say that, for given \(\{\nu,\delta\}\), there is long-lived coexistence in the population of fluctuating size and volume, if $\tau>2\langle N\rangle\approx 2\langle N\rangle_{\nu,\delta}$. The rationale for this definition is the same as in the case of  static environment, see section~\ref{Sec:StaticEnv:FixAndCoex} and \cite{cremer2009edge,he2011,reich2007,HNARM23,AHNRM23}, with the difference that the constant population is now replaced by the average stationary population size $\langle N\rangle$ that we  analytically approximate by $\langle N\rangle_{\nu,\delta}$ obtained from the $N$-PDMP according to equation~\eqref{eq:avN}. Therefore, we will say that, for given environmental parameters \(\{\nu,\delta\}\), there is {\it fast fixation} or, equivalently, {\it early extinction} of a strain, if $R$ or $S$ fixates in a time  $\tau$ that is less than $2\langle N\rangle\approx 2\langle N\rangle_{\nu,\delta}$.

Secondly, we distinguish two fluctuating environment regimes in which the fixation and long-lived coexistence probability are evaluated by distinct suitable approaches. We refer to these different behaviours as the {\it quenched} (Q) and {\it annealed} (A) regimes. Dynamic environments in the regime Q are so slow, \(\nu\ll1\), and/or so biased, \(\delta\approx \pm1\), that the environment is unlikely to experience any switch by time \(2\langle N \rangle_{\nu, \delta}\) regardless of the initial condition; whereas, in the same time, environments in regime A experience at least one environmental switch. Similar Q and A environmental regimes have been identified elsewhere, see e.g. \cite{Shnerb23,MM23}\footnote{It is worth noting that the annealed regime often corresponds to the limit $\nu\to \infty$ where the environmental noise $\xi$ self averages~\cite{Shnerb23,MM23,wienand2017evolution,wienand2018eco,taitelbaum2020population,taitelbaum2023evolutionary}. Here, the case of intermediate switching with a finite number of switches prior to fixation/extinction is qualitatively similar to the case $\nu\to \infty$, and for simplicity we here refer to it as the annealed regime A.}. The border between Q and A regimes is determined by \[\text{max}\left(1/\nu_{\pm}\right)\equiv 1/\left(\nu(1-|\delta|)\right)=2\langle N \rangle_{\nu,\delta},\] which splits the fixation-coexistence $\nu-\delta$ diagrams of figure~\ref{fig:PhaseDiagram} in two regions separated by  green/yellow lines: where $1/\left(\nu(1-|\delta|)\right)>2\langle N \rangle_{\nu,\delta}$, the  average time for an environmental switch to occur exceeds the reference time for long-lived coexistence. This means that in this region $K$ is unlikely to experience any switch by time \(2\langle N \rangle_{\nu, \delta}\), and the previous inequality hence delineates the Q regime. Here, the A regime is thus defined by $1/\left(\nu(1-|\delta|)\right)<2\langle N \rangle_{\nu,\delta}$, where, on average, $K$ switches at least once by  time  \(2\langle N \rangle_{\nu, \delta}\).

{\bf Quenched environmental regime.} In regime Q (on the left of green/yellow lines of figure~\ref{fig:PhaseDiagram}), the carrying capacity remains at initial value, $K(t)=K(0)$, at least until time \(2\langle N \rangle_{\nu,\delta}\). Since the environment is already in its quasi-stationary state, this initial value is either $K(0)=K_+$, with probability \((1+\delta)/2\), or $K(0)=K_-$, with probability \((1-\delta)/2\). Hence, the joint probability of \(R\) fixating and that fixation/extinction occurs in a time $\tau<2\langle N \rangle_{\nu, \delta}$ (fast fixation) in the regime Q, denoted by $\Phi^\text{Q}$, is 
\begin{equation}
\begin{aligned}
    \Phi^\text{Q}&\equiv \text{Prob.}(R \text{ fixation},~\tau< 2\langle N \rangle_{\nu, \delta}~|~\text{regime Q})=\\
    &=\frac{1+\delta}{2}~\phi_{K_+}\left(1-\eta_{K_+}^Q\right) + \frac{1-\delta}{2}~\phi_{K_-}\left(1-\eta_{K_-}^Q\right),
    \label{eq:QFixCoexProb}
\end{aligned}
\end{equation}
where \(\phi_N\) is the \(R\) fixation probability in a population of fixed size \(N\), given by equation~\eqref{eq:ApproxFixProb}, and
\begin{equation}
\eta_N^Q\equiv {\rm Prob.}\left(\tau>2\langle N \rangle_{\nu, \delta}\right)= {\rm exp}\left(-\frac{2\langle N \rangle_{\nu, \delta}}{\langle \tau_N\rangle}\right)
\label{eq:etaQ}
\end{equation}
is the probability of long-lived coexistence in the Q regime, starting (and staying) at a total population $N$, with mean coexistence time (MCT) $\langle\tau_N\rangle$, where $\eta_N^Q$ is defined as the environmental-dynamic counterpart of the environmental-static equation~\eqref{eq:StatCoexProb}. 

{\bf Annealed environmental regime.} In regime A (on the right of green/yellow lines of figure~\ref{fig:PhaseDiagram}), the environment switches at least once prior to time \(2\langle N \rangle_{\nu, \delta}\) regardless of the initial condition. In this regime, $N$ therefore experiences a broad range of values in sampling its quasi-stationary distribution, here approximated by $p_{\nu,\delta}(N)$ of equation~\eqref{eq:NPDMP}, before a fixation/extinction event or long-lived coexistence occurs. In the vein of~\cite{wienand2017evolution,wienand2018eco}, we thus assume that when a fixation event occurs, it arises in a population of size $N$ ponderated by the probabilistic weight $p_{\nu,\delta}(N)$. Following~\cite{MA10,AM10}, we know that the MCT  $\langle \tau_N\rangle$ in a population of constant size $N$ is the inverse of the probability flux towards the absorbing boundaries, and the fixation probability $\phi_N$ is given by the relative flux into the absorbing state $N_R=N$~\cite{AM10}. Hence, $\phi_N/\langle \tau_N\rangle$ is defined as the ``rate to \(R\) fixation'' in a population of size $N$, and similarly we define $(1-\phi_N)/\langle \tau_N\rangle$ as the ``rate to \(R\) extinction'' or, equivalently, as the ``rate to \(S\) fixation''. In regime A, we assume that any fixation/extinction events occur while $N$ is at quasi-stationarity and, by analogy with the $N$-constant case, experiences a rate to \(R\) fixation/extinction $\phi_N/\langle \tau_N\rangle$ and $(1-\phi_N)/\langle \tau_N\rangle$, respectively; but we now average these over $p_{\nu,\delta}(N)$ to account for the varying population size in fluctuating environments. We thus obtain the {\it effective rates} of \(R\) fixation/extinction in regime A, respectively given by 
 \begin{equation}
 \label{eq:effrates}
    \mathcal{T}_F = \int_{K_-}^{K_+}\frac{\phi_N}{\langle \tau_N\rangle} p_{\nu,\delta}(N)~\text{d}N\hspace{0.5cm}\text{and}\hspace{0.5cm}\mathcal{T}_E = \int_{K_-}^{K_+}\frac{1-\phi_N}{\langle \tau_N\rangle} p_{\nu,\delta}(N)~\text{d}N.
\end{equation}
Since $\langle \tau_{K_+}\rangle \gg \langle \tau_{K_-}\rangle$, the main contributions to $\mathcal{T}_{F/E}$, and hence to fixation/extinction, arise from $N\approx K_-$. With $\mathcal{T}_{F/E}$, we obtain the $R$ fixation probability and mean coexistence time, $\phi^\text{A}$ and $\langle \tau^\text{A}\rangle$, respectively, in formal analogy with the environmental-static case~\cite{MA10,AM10}:
\begin{equation}
     \phi^\text{A} = \frac{\mathcal{T}_F}{\mathcal{T}_F + \mathcal{T}_E} \hspace{0.5cm}\text{and}\hspace{0.5cm} \langle \tau^\text{A}\rangle = \frac{1}{\mathcal{T}_F + \mathcal{T}_E}.
     \label{eq:AFixProbAndMeanCoexTime}
 \end{equation}
Moreover, for a dynamic environment of mean switching rate and bias parameters \(\{\nu,\delta\}\), following the reasoning leading to equation~\eqref{eq:StatCoexProb}, the long-lived coexistence probability in regime A is 
\begin{equation}
\label{eq:etaA}
    \eta^\text{A}=\exp\left({-{\frac{2\langle N\rangle_{\nu,\delta}}{\langle \tau^\text{A}\rangle}}}\right).
\end{equation}
Results \eqref{eq:AFixProbAndMeanCoexTime} and \eqref{eq:etaA} allow us to find the joint probability of \(R\) fixation and fast fixation in a time $\tau<2\langle N \rangle_{\nu, \delta}$ in the regime A, denoted by $\Phi^\text{A}$, as
\begin{equation}
\label{eq:PhiA}
    \Phi^\text{A}\equiv\text{Prob.}(R \text{ fixation},~\tau <2\langle N \rangle_{\nu,\delta}|\text{ regime A})=\phi^\text{A}(1-\eta^\text{A}),
\end{equation}
where we have again assumed that strain fixation type and long-lived coexistence are completely uncorrelated (assumption to be assessed by means of simulations).

{\bf Crossover regime and general results.} Having obtained the fixation and coexistence probabilities in regimes Q and A, we can suitably superpose their expressions to obtain predictions applicable in the crossover regime (about the green/yellow line in figure~\ref{fig:PhaseDiagram}), as well as in regimes Q and A. This provides us with general results, that are valid for then entire range of environmental parameters $\{\nu, \delta\}$. Since the probability that no switches occur by time  $2\langle N \rangle_{\nu, \delta}$ (regime Q) is \[\Pi\equiv \exp\left[-2\langle N \rangle_{\nu,\delta}~\nu\left(1-|\delta|\right)\right],\] and the probability that at least one switch has occurred by $2\langle N \rangle_{\nu, \delta}$ (regime A) is $1-\Pi$, the overall joint probability of $R$ fixation and fast fixation is
\begin{equation}
    \Phi = \Pi \Phi^\text{Q} + (1-\Pi)\Phi^\text{A}.
    \label{eq:FinalPhi}
\end{equation}
The overall probability of long-lived coexistence is obtained by a similar superposition of $\eta^\text{Q,A}$:
\begin{equation}
    \eta = \Pi \eta^\text{Q} + (1-\Pi)\eta^\text{A}.
    \label{eq:Finaleta}
\end{equation}
Equations~\eqref{eq:FinalPhi} and \eqref{eq:Finaleta} are expressions used in the theoretical predictions of figures~\ref{fig:PhaseDiagram} and~\ref{fig:DynEnvProb}.

\subsection{Long-time eco-evolutionary dynamics in a fluctuating environment: comparison of theory and simulations}
\label{Sec:DynEnv:FixAndCoexDiagram}
In this section, we compare the results of extensive computer simulations  and theoretical predictions of section~\ref{Sec:DynEnv:Theory}, fully characterising the long-time eco-evolutionary dynamics of the population in a fluctuating environment.

\begin{figure}
\centering
\includegraphics[width=1\textwidth]{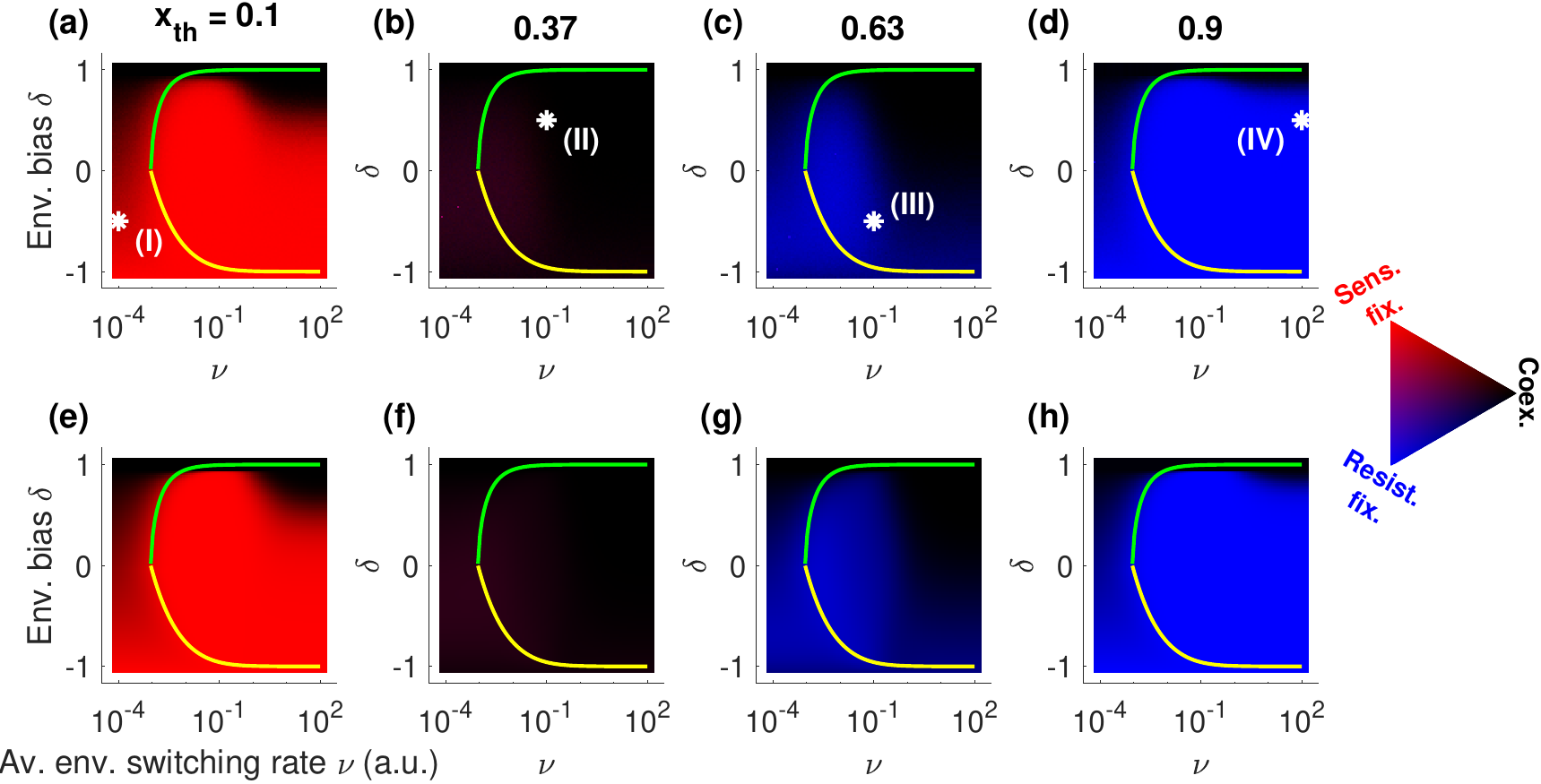}
\caption{\label{fig:PhaseDiagram}\textbf{Characterisation of the long-time 
eco-evolutionary dynamics by the fixation-coexistence diagrams.} {\bf (a)-(d)} Fixation type and fast fixation joint probability {\it in silico} for EV parameters $\nu$ (average environmental switching rate) and $\delta$ (environmental switching bias), \(s=0.1\), \(a=0.25\), \(K_-=100\),  \(K_+=1000\), and the concentration cooperation thresholds \(x_{\text{th}}=0.1,~0.37,~0.63,\) and \(0.9\). Here, $x_{\text{th}}^*\approx 0.366$, and the total population is initialised at quasi-stationarity (assumed to coincide with equation \eqref{eq:NPDMP}), with \(N_R(t=0)=x_{\text{th}}K(t=0)\) and \(N_S(0)=\left(1-x_{\text{th}}\right)K(0)\), see appendix section~\ref{Supp:Sec:sim}. Stronger blue (red) corresponds to a higher fixation probability of \(R\) (\(S\)). Darker colour indicates a higher long-coexistence probability, defined as the probability to have no fixation event by time \(2\langle N\rangle_{\nu,\delta}\), where we take twice the average total population in its stationary state (average across \(10^{3}\) realisations). The green/yellow lines separate the environmental regimes Q and A (respectively on the left and right of the lines), see section~\ref{Sec:DynEnv:Theory}. The white asterisks in (a)-(d) refer to the values of $\nu$ and $\delta$ used for each of the panel columns (I-IV) in figure~\ref{fig:Trajectories}. {\bf (e)-(h)} Theoretical fixation-coexistence diagrams: same as in panels (a)-(d) for the theoretical predictions of the fixation-coexistence joint probability given by \eqref{eq:FinalPhi} and \eqref{eq:Finaleta}, where \(2\langle N\rangle_{\nu,\delta}\) is computed as twice the average over the distribution~\eqref{eq:NPDMP}. Analytical results reproduce remarkably the features of those from simulations.}
\end{figure}

The fixation-coexistence diagrams in figure~\ref{fig:PhaseDiagram}(a)-(d) show simulation results for the probability of fast \(R\) fixation (blue), fast \(S\) fixation  (red), or long-lived coexistence (black), see appendix section~\ref{Supp:Sec:sim}, for different values of the concentration cooperation threshold $x_{\text{th}}$ when $N$ is at quasi-stationarity. The main features of these diagrams can be understood in terms of the analysis carried in section~\ref{Sec:StaticEnv:FixAndCoex} and by referring to the critical cooperation threshold value $x_{\text{th}}^*\approx [\ln\left(1-s\right)]/[\ln\left(1-a\right)]$~\footnote{In principle, a suitable environment-dynamic counterpart of $x_{\text{th}}^{*}$, given by equation~\eqref{eq:ApproxFixProb}, in a population of average size $\langle N\rangle_{\nu,\delta}$ is $x_{\text{th},\nu,\delta}^{*}\equiv\frac{\ln\left(1-s\right)}{\ln\left(1-a\right)}+\frac{1}{\langle N\rangle_{\nu,\delta} }\frac{\ln\left(\frac{s(1-a)}{a-s}\right)}{\ln\left(1-a\right)}$. However, when $\langle N\rangle\gg 1$ and $a$ is not too close to 1, as in our examples, $x_{\text{th},\nu,\delta}^{*}$ is well approximated by the leading contribution to $x_{\text{th}}^*$, and hence we can consider $x_{\text{th},\nu,\delta}^{*}\approx x_{\text{th}}^*\approx \frac{\ln\left(1-s\right)}{\ln\left(1-a\right)}$.}: when $x_{\text{th}}<x_{\text{th}}^*$ (and $x_{\text{th}}\lesssim 1/\sqrt{\langle N \rangle_{\nu,\delta}}$ for significant demographic fluctuations), the most likely outcome is the fast fixation of $S$ (i.e., early extinction of $R$); while fast fixation of $R$ (early extinction of $S$) is the most probable outcome when $x_{\text{th}}>x_{\text{th}}^*$ (and $1-x_{\text{th}}\lesssim 1/\sqrt{\langle N \rangle_{\nu,\delta}}$); and long-lived coexistence of $R$ and $S$ is expected when $x_{\text{th}}$ is close to $x_{\text{th}}^*$  (and/or when ${\rm min}(x_\text{th},1-x_\text{th})\gg 1/\sqrt{\langle N \rangle_{\nu,\delta}}$ so that demographic fluctuations are negligible). For the parameters of figure~\ref{fig:PhaseDiagram}, $x_{\text{th}}^*\approx 0.366$, and we indeed notice that most of figure~\ref{fig:PhaseDiagram}(a) appears in red, panels (c) and (d) of  figure~\ref{fig:PhaseDiagram} are mostly  coded in blue, and most of figure~\ref{fig:PhaseDiagram}(b) appears in black. 
 
\begin{figure}
\centering
\includegraphics[width=1\textwidth]{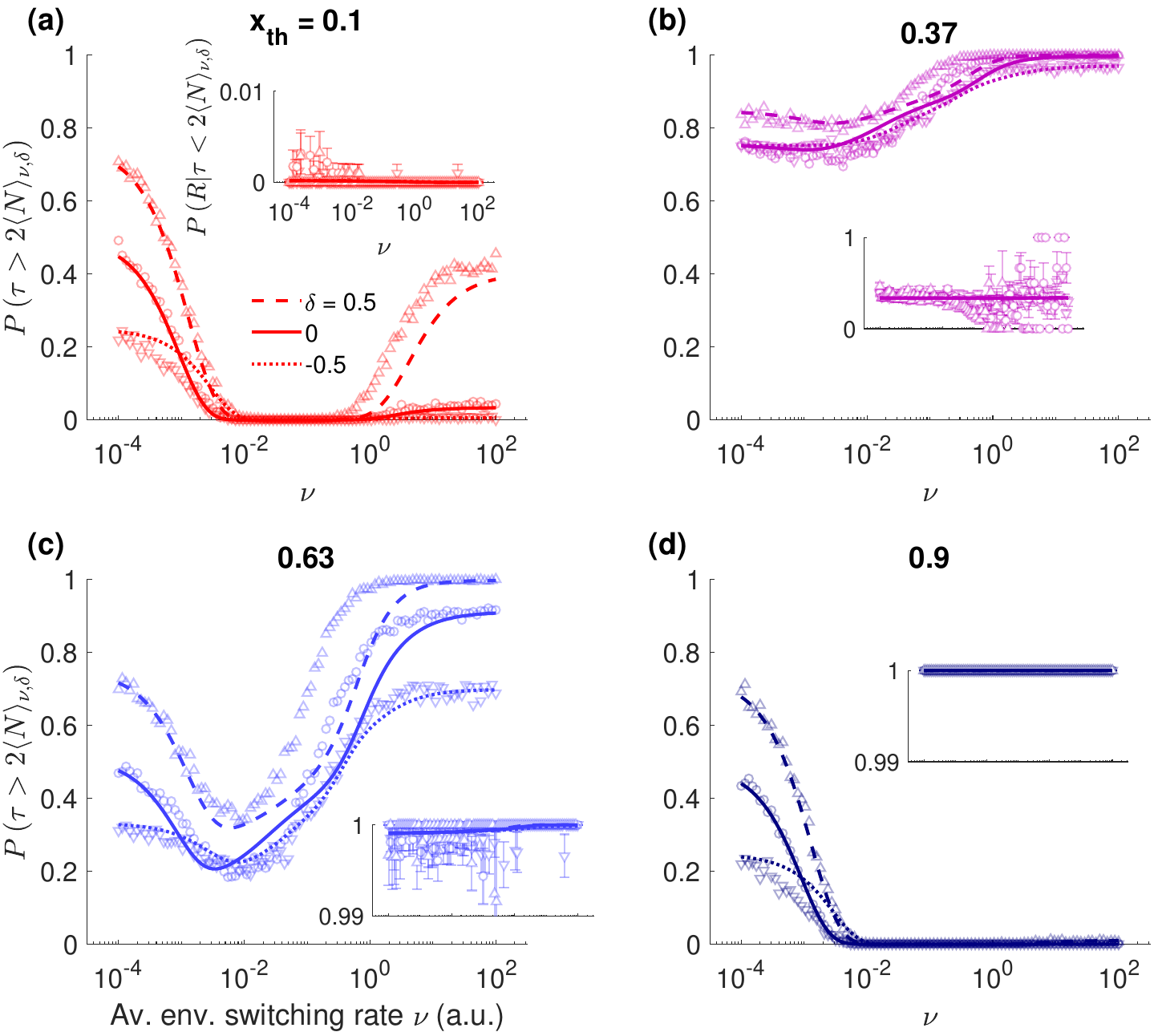}
\caption{\label{fig:DynEnvProb}
\textbf{Eco-evolutionary probabilities of long-coexistence and fast \(R\) fixation.} {\bf (a)} Long coexistence probability at quasi-stationarity \(P\left(\tau> 2\langle N\rangle_{\nu,\delta}\right)\), defined as no fixation occurring before \(2\langle N\rangle_{\nu,\delta}\), as a function of the average environmental switching rate \(\nu\) for a concentration cooperation threshold of \(x_{\text{th}}=0.1\) and different values of the environmental bias $\delta$. By \(t=2\langle N\rangle_{\nu,\delta}\), we assume that \(N\) has already reached its quasi-stationary distribution, which we approximate as equation~\eqref{eq:NPDMP} in the theoretical predictions. Dotted lines (downward triangles), solid lines (circles), and dashed lines (upward triangles) show results for \(\delta=-0.5\), \(0\), and \(0.5\), respectively. Lines are theoretical predictions from $\eta$ in equation~\eqref{eq:Finaleta}, and markers are simulation data; error bars are smaller than markers (not shown). Other parameters are: \(s=0.1\), \(a=0.25\), \(K_{-}=100\) and \(K_{+}=1000\), as in figure~\ref{fig:PhaseDiagram}. Inset: \(R\) fixation probability conditioned on fast fixation \(P\left(R|\tau< 2\langle N\rangle_{\nu,\delta}\right)\) as function of \(\nu\); lines are theoretical predictions from $\Phi/\left(1-\eta\right)$, see equations~\eqref{eq:FinalPhi}-\eqref{eq:Finaleta}; all three theoretical lines for each \(\delta\) are indistinguishable. Error bars show the binomial standard error of the mean. {\bf (b)-(d)} Same as in (a) for \(x_{\text{th}}=0.37\), \(0.63\), and \(0.9\). Note that simulation data presents larger error bars for \(\nu>10^{-1}\) in the inset of panel (c) due to the limited proportion of fast fixations in this regime.}
\end{figure}

We notice however that in the presence of EV each panel of figure~\ref{fig:PhaseDiagram} is coded in two colours, indicating a phase of fast fixation/extinction and another of long-lived coexistence. In fact, due to EV, long-lived coexistence is possible even when fast fixation of $S$ and $R$ is most likely. In figure~\ref{fig:PhaseDiagram}(a),(d) this leads to the red and blue dominated diagrams to also contain a dark/black area for $\delta$  close to $1$ (bias of $K$ towards $K_+$). This can be qualitatively explained by noting that, as  $K$ is biased towards $K_+$, the population size is essentially constant and large, $N\approx K_+$, implying that demographic fluctuations are weak, which prevents fast fixation of either strain. Similarly, EV can allow for fast fixation even when long-lived coexistence is expected, as found in figure~\ref{fig:PhaseDiagram}(b), where the black-dominated diagram (due to $x_{th}\approx x^*_{th}$) contains a coloured ``cloud''. Here, fast fixation events occur for low values of $\nu$ and $\delta\leq 0$, when $N\approx K_-$ for long periods, and the small population size ($K_-=100$) enforces large demographic fluctuations that facilitate early extinction of either $R$ or $S$, see figure~\ref{fig:DynEnvProb}.

In general, the colour-coding of the fixation-coexistence $\nu-\delta$ diagrams of figure~\ref{fig:PhaseDiagram} can be qualitatively understood by noting on the one hand that, as $\delta$ is increased from $-1$ to $1$ (at fixed $\nu$), the EV bias moves from $K_-$ (small community) to $K_+$ (large population), and hence favours long-lived coexistence. On the other hand, increasing the environmental switching rate from slow (\(\nu\ll1\)) to intermediate (\(\nu\sim1\)) hinders strain coexistence since, in intermediate switching, irrespective of the starting environment, the microbial community experiences long population bottlenecks (\(N\approx K_{-}\)) that enhance demographic fluctuations and thus favour fast fixation. However, further increasing the rate to fast switching (\(\nu\gg1\)) increases the probability of long-lasting coexistence. Namely, the duration of population bottlenecks is reduced, and the community effectively experiences a higher carrying capacity \(\mathcal{K}>K_{-}\), with lower demographic fluctuations and enhanced coexistence. Note that, for extreme values of \(x_{\text{th}}\), which are closer to extinction/fixation of one strain (\(x=0,1\)), the latter does not hold, as weak fluctuations suffice to prevent coexistence.

In the aim of understanding the evolution of AMR,  we are particularly interested in finding the conditions favouring the early extinction of $R$, and hence the fast fixation of $S$, since this corresponds to the most favourable environmental conditions for the {\it early eradication of AMR}~\cite{HNARM23}. Here, the best conditions for AMR eradication appear as red areas in the diagrams of figure~\ref{fig:PhaseDiagram}: these are low values of $x_{\text{th}}$ and $\delta$ not too close to $1$, see figure~\ref{fig:PhaseDiagram}(a). Moreover, as a signature of the influence of EV, discussed above, we find that AMR eradication is also possible for intermediate values of $x_{\text{th}}\gtrsim x_{\text{th}}^*$, see reddish area in figure~\ref{fig:PhaseDiagram}(b) and inset of figure~\ref{fig:DynEnvProb}(b).

In summary, a high EV bias $\delta$ (at fixed $\nu$) and a moderate cooperation threshold (\(x_{\text{th}}\approx x^*_{\text{th}}\)) favour long-lived coexistence; while intermediate EV switching rates $\nu$ (at fixed $\delta$), and more extreme cooperation thresholds, favour fast fixation of either \(R\) (for \(x_{\text{th}}\gg x^*_{th}\)) or \(S\) (when \(x_{\text{th}}\ll x^*_{th}\)). This qualitatively explains the boundaries between the coloured and black regions in the fixation-coexistence diagrams of figure~\ref{fig:PhaseDiagram}; and, notably, it captures the long-lived coexistence in the top/right-half and fast $R$ fixation in the bottom/left-half of figure~\ref{fig:PhaseDiagram}(c). Critically, the location of the boundaries between the phases of  long-lived and fast fixation are accurately predicted by the  theory presented in section~\ref{Sec:DynEnv:Theory}, as shown in figure~\ref{fig:PhaseDiagram}(e)-(h) where the fixation-coexistence diagrams are reproduced remarkably well by the predictions of \eqref{eq:FinalPhi} and \eqref{eq:Finaleta}.

Figure~\ref{fig:DynEnvProb} provides a more quantitative comparison between the fixation-coexistence diagrams obtained from simulations in figure~\ref{fig:PhaseDiagram} (top row) and from the theory of section~\ref{Sec:DynEnv:Theory} (bottom row), for three different values of environmental bias (high/low \(\delta\) and $\delta=0$). In each panel we find that the $\nu$-dependence of the long-lived coexistence probability, that is, the probability that $\tau>2\langle N\rangle_{\nu,\delta}$ in simulations, is well captured by the theoretical prediction of \(\eta\) given by~\eqref{eq:Finaleta}. The agreement between simulations and \(\eta\) is quantitatively remarkable at low and high values of $\nu$. For intermediate values of $\nu$, the predictions of $\eta$ are able to reproduce the main features of the simulation data, including the non-monotonic behaviour arising in figure~\ref{fig:DynEnvProb}(a),(c). The values of $\nu$ for which the probability of long-lived coexistence is low correspond to regions coloured in red/blue in figure~\ref{fig:PhaseDiagram}. The main deviations between theory and simulations arise in the regime of intermediate $\nu$ and high $\delta$. We attribute these deviations to the assumptions made in the annealed regime (approximation of the actual fixation and extinction rates), leading to an underestimate of the long-lived coexistence probability. The inset of each panel of figure~\ref{fig:DynEnvProb} illustrates the \(R\) fixation probability, conditioned on fast fixation (\(\tau<2\langle N\rangle_{\nu,\delta}\)), which determines the relative red-to-blue colour levels in figure~\ref{fig:PhaseDiagram}. Consistently with the fixation-coexistence diagrams of figure~\ref{fig:PhaseDiagram}, fast \(R\) fixation  is found to be essentially independent of switching rate \(\nu\) and bias \(\delta\), a feature that is well reproduced by the theoretical predictions (lines in figure~\ref{fig:DynEnvProb}), obtained from equations~\eqref{eq:FinalPhi} and \eqref{eq:Finaleta} as \[\Phi/\left(1-\eta\right)\equiv\text{P}\left(R|\tau<2\langle N\rangle_{\nu,\delta}\right).\] The inset of figure~\ref{fig:DynEnvProb}(b) indicates that the probability of fast $S$ fixation is more likely than fast 
\(R\) fixation (whose probability is less than $1/2$),  which explains that reddish ``coloured cloud'' in figure~\ref{fig:PhaseDiagram}(b).

\section{Conclusions}
\label{Sec:Conclusion}
Microorganisms live in endlessly changing environments, and experience conditions that often fluctuate between mild and harsh. This environmental variability (EV) often affects the amount of nutrients and toxins in a community, and thus shapes the eco-evolutionary dynamics of microbial communities. EV often crucially influences the ability of species to coexist and cooperate, a feature that is key to better understand the evolution of antimicrobial resistance (AMR), whose rise is a global societal threat~\cite{oneill2016tackling,Lancet2022}. In this context, a central question concerns how EV affects the coexistence of microbes resistant and sensitive to antimicrobial drugs. It is well established that certain resistant cells are able to inactivate antimicrobials and, under certain conditions, share the protection against the drugs with the entire microbial community. This form of AMR can be regarded as cooperative behaviour. In this work, we have studied an idealised model of cooperative AMR where sensitive and resistant cells compete for the same resources in a microbial community of fluctuating size, in the presence of a biostatic drug, and subject to demographic fluctuations and environmental variability (EV). The latter is modelled by a binary carrying capacity randomly switching between high and low values (mild and harsh  conditions, respectively), and is responsible for the time-variation of the community volume and size: the total amount of nutrients and toxin in the community fluctuates, but their concentration is kept fixed (time-changing volume). When the concentration of resistant microbes exceeds a fixed {\it concentration cooperation threshold}, their protection, provided by drug-inactivating enzymes, is shared with sensitive cells at no metabolic cost. However, when the resistant cells concentration is below the cooperation threshold, only resistant microbes benefit from the protection offered by the enzymes. In this setting, the evolutionary dynamics of AMR can be viewed as the eco-evolutionary dynamics of a public good game in a fluctuating environment for a  microbial community of time-varying volume. In a large population under static environment, fluctuations are negligible and both strains always coexist. Moreover, in the absence of EV, the fate of a community of finite size is chiefly determined by the value of the  concentration  cooperation threshold \(x_{\text{th}}\) relative to a critical value \(x^*_{\text{th}}\) that we have analytically determined. In particular, when the concentration  cooperation threshold is close to 0, which means that resistant microbes are very efficient at generating drug-inactivating enzymes, resistant cells are likely to become extinct. On the other hand, when resistant microbes are inefficient at inactivating the drug, the cooperation threshold is close to 1, and the resistant strain takes over the entire microbial population. Otherwise, when the cooperation threshold is close to the critical value \(x^*_{\text{th}}\), a long-lived coexistence of the strains is expected to set in.

Here, we show that this picture is drastically altered by the joint effect of EV and demographic fluctuations in a dynamic environment. Using computational and analytical tools, we show that, for a range of different values of the cooperation threshold, both coexistence and fixation of one of the strains is possible. By computing the fixation-coexistence diagrams of the model, we have determined the nontrivial environmental conditions (switching rate and bias) separating the phase of dominance/fixation of one strain from that of long-lived coexistence of both species. We have therefore shown that, in the presence of EV, the long-time eco-evolutionary dynamics cannot be predicted only from the value of the concentration cooperation threshold \(x_{\text{th}}\): in the presence of EV, long-lived coexistence of the strains is possible even when fast fixation/extinction is likely (for \(x_{\text{th}}\) close to 0 or 1), and there can be early extinction/fixation of a strain even when long-lived coexistence is expected (for \(x_{\text{th}}\) close to \(x^*_{\text{th}}\)). Our results show that increasing the EV bias (at fixed switching rate) favours long-lived coexistence, while intermediate EV switching rates (at fixed  EV bias) favour fast fixation of one of the strains. We have rationalised our findings by devising an analytical approach, built on the combination of suitable quenched and annealed averaging procedures in different EV regimes, allowing us to reproduce qualitatively and quantitatively the diagrams and fixation-coexistence properties obtained from simulations. This has allowed us to analytically characterise and reproduce the full two-phase fixation-coexistence diagrams, and to obtain the long-lived coexistence probability of the strains, as well as the conditional probability that resistant cells rapidly take over the microbial population, as a function of the environmental variation rate and cooperation threshold. Importantly in the context of modelling the evolution of AMR, our findings allow us to identify the most favourable environmental conditions for the {\it early eradication of AMR}, which correspond to an early extinction of resistant cells.

It is worth noticing that in~\cite{HNARM23} we considered a related model with  environmental fluctuations arising at {\it constant volume}, i.e. a form of EV with time-varying concentration of nutrients and toxins, which is characterised by fixation-coexistence diagrams of three phases (fixation of each strain and long-lived coexistence), and a fluctuation-driven AMR eradication mechanism. Hence, the findings reported in this work complement those of~\cite{HNARM23} and give us a broader perspective on the joint influence of environmental and demographic fluctuations on the evolution of cooperative AMR, paving the way for numerous possible applications. We also believe that the analytical methods devised in this work, and their generalisations, shall be applicable to describe the eco-evolutionary dynamics of a broad class of systems.

\vspace{1cm}

\section*{Data accessibility}
Simulation data and codes for all figures are electronically available from the University of Leeds Data Repository. DOI: \href{https://doi.org/10.5518/1462}{10.5518/1462}.

\section*{Author Contributions}
{\bf Llu\'is Hern\'andez-Navarro:} Conceptualisation (lead), Methodology (lead), Formal Analysis, Software, Writing - Original Draft, Writing - Review \& Editing, Visualisation, Investigation, Validation. {\bf Matthew Asker:} Formal Analysis, Software, Writing - Original Draft, Writing - Review \& Editing, Visualisation, Investigation, Validation. {\bf Mauro Mobilia:} Conceptualisation, Methodology, Writing - Original Draft, Writing - Review \& Editing, Visualisation, Supervision, Project administration, Funding acquisition.

Contributor roles taxonomy by CRediT~\cite{brand2015beyond}.

\section*{Competing interests}
We declare we have no competing interests.

\section*{Funding}
L.H.N. and M.M. gratefully acknowledge funding from the U.K. Engineering and Physical Sciences Research Council (EPSRC) under the grant No. EP/V014439/1 for the project `DMS-EPSRC Eco-Evolutionary Dynamics of Fluctuating Populations’. The support of a Ph.D. scholarship to M.A. by the EPSRC  grant No. EP/T517860/1  is also thankfully acknowledged. 

\section*{Acknowledgements}
We dedicate this paper to Uwe T\"auber for his numerous influential contributions to statistical physics. 
We are also grateful to K. Distefano, J. Jim\'enez, S. Mu\~{n}oz Montero, M. Pleimling, A. M. Rucklidge, and M. Swailem for useful discussions. This work was undertaken on ARC4, part of the High Performance Computing facilities at the University of Leeds, UK.


\newpage

\setcounter{figure}{0}
\renewcommand{\figurename}{Figure}
\renewcommand{\thefigure}{S\arabic{figure}}
\setcounter{equation}{0}
\renewcommand{\theequation}{S\arabic{equation}}

\appendix
\title{{\huge\textbf{Appendix}}}
\section{Simulation methods}\label{Supp:Sec:sim}
We have studied the full eco-evolutionary dynamics 
of the model by carrying out extensive \textit{exact}
stochastic simulations following the Next Reaction Method of references~\cite{Gibson00,Anderson07}, and proceeding as in~\cite{AHNRM23,HNARM23}, where details and code are presented in the supplemental material~\cite{SM_HNARM23}. The simulation data and codes that we have generated are electronically available, alongside comments, in~\cite{data}.

The system is always initialised at stationarity, and allowed to evolve according to the reactions and rates defined in section~\ref{Sec:MethMod}, which are summarised in the master equation~\eqref{eq:ME}. The starting populations are always set at \(N_R(t=0)=x_{\text{th}}N(0)\) and \(N_S(t=0)=\left(1-x_{\text{th}}\right)N(0)\). For simplicity, the initial total population follows \(N(0)=K(0)\), instead of sampling from the approximate quasi-stationary distribution~\eqref{eq:NPDMP}, since the observed results are virtually the same. All simulation data points are averaged over $10^3$ realisations, except in figures~\ref{Fig:Sketch}(c)-(d) and~\ref{fig:Trajectories}, where single trajectories are shown. In figure~\ref{fig:DynEnvProb}, error bars are estimated as the binomial standard error of the mean \(\sqrt{\hat{p}\left(1-\hat{p}\right)/n}\), where \(\hat{p}\) is the estimated average probability and \(n\) is the number of realisations considered. For the main plot in each panel, error bars are smaller than marker size, and thus not shown. The larger error bars observed in the inset of panel (b) at \(\nu>10^{-1}\) arise from the low number of realisations showing fast fixation (\(\tau<2\langle N\rangle_{\nu.\delta}\)) in this regime due to the high probability of long coexistence.

\section{Fixation probability}
\label{Supp:Sec:ExactFixProb}
For the Moran model and a population of constant size $N$, the exact general expression of the $R$ fixation probability depending on the initial number of \(R\) cells is~\cite{Gardiner, VanKampen, Ewens, antal2006fixation, traulsen2009stochastic}
\begin{equation}
    \phi_N\left(N_R^0 \right)=\frac{1+\sum_{k=1}^{N_R^0-1}\prod_{i=1}^{k}\gamma\left(i,N\right)}{1+\sum_{k=1}^{N-1}\prod_{i=1}^{k}\gamma\left(i,N\right)},~\text{with}~\gamma\left(N_R,N\right)\equiv\frac{\widetilde{T}^-_R\left(N_R,N\right)}{\widetilde{T}^+_R\left(N_R,N\right)}~\text{and}~N_R^0=1, 2,..., N,
    \label{SuppEq:GenFixProb}
\end{equation}
where \(\widetilde{T}^{\pm}_R\left(N_R,N\right)\) are defined in the main equation~\eqref{eq:morantransrates} and the function \(\gamma\left(N_R,N\right)\) fully determines the fixation probability. When $N$ is sufficiently large, the initial fraction of resistant cells $x_0\equiv N_R^0/N$ can be approximated to a continuous variable. The fixation probability can thus be regarded as a continuous function of $x_0$, and it is convenient to write $\phi_N\left(N_R^0=Nx_0\right)$ as $\phi_N\left(x_0\right)$.

As discussed in main section~\ref{Sec:StaticEnv:FixAndCoex}, when $x_0$ is not too close to $0$ or $1$, we can assume that the fixation/extinction of a strain occurs from \(x_{0}=x_{\text{th}}\), i.e., \(\phi_N(x_0)\simeq\phi_N\left(x_0=x_{\text{th}}\right)\). In this case, the fixation probability can be simplified to the following exact expression~\cite{HNARM23}:
\begin{equation}
    \phi_N(x_{\text{th}})=\frac{1-\left(\frac{1-a}{1-s}\right)^{x_{\text{th}}N}}
    {1-\left(\frac{1-a}{1-s}\right)^{x_{\text{th}}N}+\frac{a-s}{s(1-a)}\left(\frac{1-a}{1-s}\right)^{x_{\text{th}}N}\left[\left(\frac{1}{1-s}\right)^{(1-x_{\text{th}})N}-1\right]}.
    \label{SuppEq:FixProb}
\end{equation}
The compact approximation of main equation~\eqref{eq:ApproxFixProb} has been obtained by neglecting \((1-a)^{x_{\text{th}}N}/(1-s)^{x_{\text{th}}N}\ll 1\) and \((1-s)^{(1-x_{\text{th}})N}\ll1\).

\section{Mean Coexistence Time (MCT)}
\label{Supp:SubSec:ExactMCT}
The exact expression of the MCT for the Moran model in a fixed population of fixed size $N$ is well-known~\cite{Gardiner, VanKampen, Ewens, antal2006fixation, traulsen2009stochastic}, and can here be written as
\begin{equation}
    \langle \tau_N\left(N^0_R\right)\rangle =
    \phi_N(N^0_R)\sum_{k=N_R^0}^{N-1}\sum_{n=1}^{k}\frac{\prod_{m=n+1}^{k}\gamma\left(m,N\right)}{\widetilde{T}^+_R\left(n,N\right)}-\left[1-\phi_N(N^0_R)\right]\sum_{k=1}^{N_R^0-1}\sum_{n=1}^{k}\frac{\prod_{m=n+1}^{k}\gamma\left(m,N\right)}{\widetilde{T}^+_R\left(n,N\right)},
    \label{SuppEq:GenMeanAbsTime}
\end{equation}
where the subscript indicates that the population size $N$ is fixed (here $N=K_0$). The function \(\phi_N(N^0_R)\) is the \(R\) fixation probability given by equation~\eqref{SuppEq:GenFixProb}, where the functions \(\widetilde{T}^+_R\left(N_{R},N\right)\) and \(\gamma\left(N_{R},N\right)\) are also defined.

Supported by the assumption \(x_0\simeq x_{\text{th}}\) of section~\ref{Sec:StaticEnv:FixAndCoex}, we focus on the case \(N_R^0=N_{\text{th}}\equiv x_{\text{th}}N\) to simplify equation~\eqref{SuppEq:GenMeanAbsTime}. Note that, in this case, the last double summation has powers of \(\gamma=(1-a)/(1-s)<1\), while the first double summation also has additional summands with high powers of \(\gamma=1/(1-s)>1\). Hence, the last double summation is negligible with respect to the first one. Furthermore, if we split the inner summation of the first double summation at \(n=N_{\text{th}}\) we obtain
\begin{equation}
    \begin{aligned}
    \langle\tau_N\left(x_0\right)\rangle\simeq \langle\tau_N\left(x_{\text{th}}\right)\rangle\simeq
    \phi_N\left(x_{\text{th}}\right)\sum_{k=N_{\text{th}}}^{N-1}\left(\sum_{n=1}^{N_{\text{th}}-1}\frac{\prod_{m=n+1}^{k}\gamma\left(m/N\right)}{\widetilde{T}^+_R\left(n/N\right)}+\sum_{n=N_{\text{th}}}^{k}\frac{\prod_{m=n+1}^{k}\gamma\left(m/N\right)}{\widetilde{T}^+_R\left(n/N\right)}\right).
    \end{aligned}
    \label{ApproxMAT_OLD2}
\end{equation}
According to the definition in section~\ref{Sec:StaticEnv:FixAndCoex}, each summand reads \[\frac{\prod_{m=n+1}^{k}\gamma\left(m/N\right)}{\widetilde{T}^+_R\left(n/N\right)}=\frac{(1-a)^{\left[\frac{\left(N_{\text{th}}-1+k\right)-\left|N_{\text{th}}-1-k\right|}{2}-\frac{\left(N_{\text{th}}-1+n\right)-\left|N_{\text{th}}-1-n\right|}{2}\right]}}{(1-s)^{k-n}}\cdot\left[\frac{\frac{1-s}{N-n}+\frac{1-a\cdot\theta\left(N_{\text{th}}-n\right)}{n}}{1-s}\right].\]
And thus, 
\begin{equation}
    \begin{aligned}
    \langle\tau_N\left(x_{\text{th}}\right)\rangle\simeq
    \frac{\phi_N\left(x_{\text{th}}\right)}{1-s}\sum_{k=N_{\text{th}}}^{N-1}\left[\left(\frac{1}{1-s}\right)^k\left((1-a)^{N_{\text{th}}-1}\sum_{n=1}^{N_{\text{th}}-1}\left(\frac{1-s}{1-a}\right)^n\left[\frac{1-s}{N-n}+\frac{1-a}{n}\right]\right.\right.\\
    \left.\left.+\sum_{n=N_{\text{th}}}^{k}(1-s)^n\left[\frac{1-s}{N-n}+\frac{1}{n}\right]\right)\right].
    \end{aligned}
    \label{ApproxMAT_OLD3}
\end{equation}
We observe that the first summation over \(n\) in the previous equation is independent of \(k\), and therefore we can take it out as a common factor. Regarding the second summation over \(n\), note that only those sums with highest upper index \(k\) are significant because they are weighted by the prefactor \((1-s)^{-k}\gg1\); furthermore, note that the summands' magnitude rapidly decreases with \(n\). By combining these two features, we realise that we can substitute the upper index of the summation \(k\) by \(N-1\) without incurring in a significant error, making the summation independent of \(k\), and taking it out as a common factor as well. Hence, we can now compute the remaining summation over \(k\) as a geometric progression. The above considerations yield
\begin{equation}
    \begin{aligned}
    \langle\tau_N\left(x_{\text{th}}\right)\rangle\simeq
    \phi_N\left(x_{\text{th}}\right)\frac{\left(\frac{1}{1-s}\right)^{N-N_{\text{th}}}-1}{s(1-s)^{N_{\text{th}}}}\left((1-a)^{N_{\text{th}}-1}\sum_{n=1}^{N_{\text{th}}-1}\left(\frac{1-s}{1-a}\right)^n\left[\frac{1-s}{N-n}+\frac{1-a}{n}\right]\right.\\
    \left.+\sum_{n=N_{\text{th}}}^{N-1}(1-s)^n\left[\frac{1-s}{N-n}+\frac{1}{n}\right]\right).
    \end{aligned}
    \label{ApproxMAT_OLD4}
\end{equation}
In the first summation over \(n\), each summand is weighted by an exponentially increasing factor \(((1-s)/(1-a))^{n}\gg1\). Hence, only the last terms with highest \(n\) are significant. For these summands we have \(n\lesssim N_{\text{th}}\), and the fractions \((1-s)/(N-n)\) and \((1-a)/n\) evolve very slowly with \(n\) compared to the exponential factor. Therefore, we can approximate \(n\) in the previous fractions by \(N_{\text{th}}\), and take them as a common factor out of the summation without adding a significant error. As for the second summation, we can also substitute \(n\) by \(N_{\text{th}}\) in the fractions and take them out as a common factor. In this case, the rationale is that the summands with lowest \(n\) (which corresponds to \(n\gtrsim N_{\text{th}}\)) are the only significant terms due to the factor \((1-s)^n\ll1\). Finally, we can compute the two remaining geometric series over \(n\), and simplify the expression to that of main equation~(\ref{eq:ApproxMAT}).

\section{Numerical marginalisation of an arbitrary function over the {\it N}-PDMP}
\label{Supp:SubSec:NPDMPmarginalisation}
In section~\ref{Sec:DynEnv:FixCoexQSCD}, we show the stationary probability density \eqref{eq:NPDMP} of the $N$-PDMP $p_{\nu,\delta}(N)$, which provides a useful proxy for the actual quasi-stationary distribution of the population size. Our theoretical analysis requires the marginalisation, and hence the integration, over $p_{\nu,\delta}(N)$, see equations~\eqref{eq:avN} and~\eqref{eq:effrates}. Here, we describe how to efficiently integrate over $p_{\nu,\delta}(N)$ numerically.

The marginal stationary probability density function  of \(N\)-PDMP  is~\cite{horsthemke1984noise,HL06,Bena2006,Ridolfi11,wienand2017evolution,wienand2018eco,taitelbaum2020population,AHNRM23}
\begin{equation}
    \begin{aligned}
    &p_{\nu,\delta}(N)=\mathcal{Z}\cdot\frac{\left(1-\frac{K_-}{N}\right)^{\nu_--1}\left(\frac{K_+}{N}-1\right)^{\nu_+-1}}{N^{2}},\text{~with the normalisation constant} \\
    &\mathcal{Z}\equiv\frac{(K_+)^{\nu_-}(K_-)^{\nu_+}\Gamma\left(\nu_++\nu_-\right)}{\left(K_+-K_-\right)^{\nu_++\nu_--1}\Gamma\left(\nu_+\right)\Gamma\left(\nu_-\right)},
    \end{aligned}
    \label{PDMP2_AnnexB}
\end{equation}
where $\Gamma(x)\equiv \int_0^{\infty} t^{x-1}e^{-t}~dt$ denotes the standard gamma function. The  arrangement of this equation~\eqref{PDMP2_AnnexB} is convenient to avoid numerical limitations, as it minimises the magnitudes of the bases of the exponents. Numerical limitations in the above equation for \(\mathcal{Z}\) at too high \(K_{\pm}\) can be sidestepped in the standard way, i.e. by directly computing the normalisation constant as the inverse of \(\int^{K_+}_{K_-}p_{\nu,\delta}(N)/\mathcal{Z}~dN\).

The general form of the integration that we want to compute numerically in an efficient manner is \[\int^{K_+}_{K_-}f(N)p_{\nu,\delta}(N) dN,\] where \(f(N)\) can be any arbitrary function that is once-differentiable in the closed domain \(N\in [K_{-},K_{+}]\). This numerical computation can be ill-posed due to singularities at the limits, an issue stemming from the factors \(\left|N-K_\pm\right|^{\nu_\pm-1}\) in the expression of \(p_{\nu,\delta}(N)\) when \(\nu_\pm<1\). To sidestep these numerical issues, we split the integration domain over three subdomains: \([K_{-},K_{-}+1)\), \([K_{-}+1,K_{+}-1]\), and \((K_{+}-1,K_{+}]\); and we then integrate by parts the first and/or last subdomains to avoid any singularities. This is, first, we integrate the diverging factor \(\left(N-K_{-}\right)^{\nu_{-}-1}\) in the first subdomain if \(\nu_{-}<1\), and/or the diverging factor \(\left(K_{+}-N\right)^{\nu_{+}-1}\) in the last subdomain if \(\nu_{+}<1\); and, second, we derive the remainder as it is standard when integrating by parts. We thus obtain
\begin{equation}
\begin{aligned}
    &\int^{K_+}_{K_-}f(N)p_{\nu,\delta}(N)dN=\\
    &f(N)\cdot p_{\nu,\delta}(N)\cdot\frac{N-K_-}{\nu_-}\bigg]^{K_-+1}_{K_-}-\int^{K_-+1}_{K_-}\frac{f'(N)-f(N)\left(\frac{\nu_+-1}{K_+-N}+\frac{\nu_++\nu_-}{N}\right)}{\nu_-}\left(N-K_-\right)p_{\nu,\delta}(N)dN\\
    &+\int^{K_+-1}_{K_-+1}f(N)p_{\nu,\delta}(N)dN+\\
    &-f(N)\cdot p_{\nu,\delta}(N)\cdot\frac{K_+-N}{\nu_+}\bigg]^{K_+}_{K_+-1}+\int^{K_+}_{K_+-1}\frac{f'(N)+f(N)\left(\frac{\nu_--1}{N-K_-}-\frac{\nu_++\nu_-}{N}\right)}{\nu_+}\left(K_+-N\right)p_{\nu,\delta}(N)dN,
\end{aligned}
\end{equation}
where now \(p_{\nu,\delta}(N)\) is always multiplied by either \((N-K_{-})\) or \((K_{+}-N)\) when \(N=K_{-}\) or \(K_{+}\), respectively. Crucially, this transforms any singularities when \(\nu_{\pm}<1\) (for \(N=K_{\pm}\)) into removable singularities. Taking into account vanishing terms, this simplifies to
\begin{equation}
\begin{aligned}
    \int^{K_+}_{K_-}f(N)p_{\nu,\delta}(N)dN=&&-\int^{K_-+1}_{K_-}\frac{f'(N)-f(N)\left(\frac{\nu_+-1}{K_+-N}+\frac{\nu_++\nu_-}{N}\right)}{\nu_-}\left(N-K_-\right)p_{\nu,\delta}(N)dN\\
    &+&\frac{f\left(K_-+1\right)p_{\nu,\delta}\left(K_-+1\right)}{\nu_-}+\int^{K_+-1}_{K_-+1}f(N)p_{\nu,\delta}(N)dN+\frac{f\left(K_+-1\right)p_{\nu,\delta}\left(K_+-1\right)}{\nu_+}\\
    &+&\int^{K_+}_{K_+-1}\frac{f'(N)+f(N)\left(\frac{\nu_--1}{N-K_-}-\frac{\nu_++\nu_-}{N}\right)}{\nu_+}\left(K_+-N\right)p_{\nu,\delta}(N)dN,
\end{aligned}
\end{equation}
where \(f'(N)\) can be computed analytically or by standard numerical methods, if needed. Therefore, we have finally obtained an exact expression containing well-posed integrals that can be computed numerically and efficiently.

\end{document}